\documentclass[preprintnumbers,onecolumn, floatfix, preprintnumbers, amsmath, amssymb, superscriptaddress]{revtex4}

\usepackage[colorlinks,linkcolor=red,urlcolor=blue,citecolor=blue]{hyperref}
\usepackage{amssymb, amsmath, bm, dcolumn, epsf, graphicx, latexsym, mathbbol, slashed, mathrsfs, comment, multirow}

\newcommand{\lambdabbar}{\lambda\kern-0.5em\raise0.5ex\hbox{--}}

\newcommand{\dd}[0]{\mathrm{d}}
\allowdisplaybreaks

\begin{document}

\title{Gravitational Lensing of Gravitational Waves: Towards a Higher-order Geometric-optics Approach}

\author{Zhao Li}
\email{lz111301@mail.ustc.edu.cn}
\affiliation{Department of Astronomy, Peking University, Beijing, 100871, China}

\author{Shaoqi Hou}
\affiliation{School of Physics and Technology, Wuhan University, Wuhan, Hubei 430072, China}


\author{Wen Zhao}
\email{wzhao7@ustc.edu.cn}
\affiliation{Department of Astronomy, University of Science and Technology of China, Hefei, Anhui 230026, China, and\\ School of Astronomy and Space Science, University of Science and Technology of China, Hefei, Anhui 230026, China}
\affiliation{College of Physics, Guizhou University, Guiyang, Guizhou 550025, China}

\begin{abstract}
In this work, we study the gravitational lensing of gravitational waves (GWs) by extending the geometric-optics approximation to higher order. With the help of the Newman-Penrose formalism, we reexpress the GW propagation equations as a series of scalar equations and present explicit expressions for the Weyl scalars that describe the GW polarizations. By combining the approaches of solving geodesic deviation and transport equations, we construct a solvable system of equations that describes the evolution of GW polarization along null geodesics. This framework fills the gap left by the leading-order geometric optics and the Kirchhoff diffraction integral, neither of which captures the polarization characteristics of GWs during the lensing process. This work applies the above framework to a Schwarzschild lensing configuration. Through a rigorous theoretical formulation and detailed numerical analysis, our results reveal the emergence of apparent vector and scalar modes in lensed GW signals, which originate from the smearing of the polarization plane and distortion of the wavefront and do not represent genuine dynamical degrees of freedom but rather arise as the propagation effects imposed by gravitational lensing.

\end{abstract}

\maketitle

\section{\label{Introduction}Introduction}

Gravitational lensing has emerged as a powerful probe at the intersection of gravitational theory, astrophysics, and modern cosmology \cite{Wambsganss_1998,Bartelmann_2010,Dodelson_2017,Refregier_2003}. In analogy with electromagnetic waves, gravitational waves (GWs) propagating through an inhomogeneous spacetime are deflected, magnified, and time-delayed by intervening matter distributions, a phenomenon referred to as GW lensing \cite{Maggiore_2008}. Since the first direct detection of GWs by the LIGO/Virgo Collaboration \cite{GW150914}, the possibility of observing lensed GW signals has attracted growing attention. Detecting such lensing signatures in GW data would provide unique information about the lens population \cite{Cao_2022,Lo_2023} and the nature and substructures of dark matter \cite{Liao_2018,Liao_2020,Guo_2022,Cheung_2024,Jana_2025}. It would offer independent constraints on modified gravity and cosmological parameters \cite{Yang_2019,Li_2019,Wang_2024,Cao_2012,Chen_2026,Mukherjee_2020,Hou_2020,Hou_2021,Fan_2017}. From the observational perspective, several searches for lensed GW events have been carried out using the GW transient catalogs from the LIGO/Virgo/KAGRA collaboration, including searching for the repeated signals from strong lensing and searching for the lens-induced waveform distortions \cite{Haris_2018,Hannuksela_2019,McIsaac_2020,Liu_2021,LIGO_2021,Abbott_2024}. While no unambiguous detection has yet been confirmed, the event GW231123 \cite{Abac_2025} has been reported to show a preference for a lensed-signal interpretation in some analyses \cite{LVK_lensing,Chan_2025_b,Goyal_2025,Chakraborty_2025,Shan_2025_b,Wang_2026}. Nevertheless, the increasing sensitivity of current and future detectors suggests that lensed GW events may soon be routinely observed \cite{Yang_2021,Gao_2022,Lin_2023,Piorkowska_2013,Juan_2025,Yuan_2026,Barsode_2026}.

As a central ingredient in GW data analysis, matched filtering technique relies on accurate theoretical templates to extract weak signals from noisy detector data and to infer source parameters. Inaccurate or incomplete templates may lead to biased parameter estimation or even missed detections \cite{Canitrot_2001,Lee_2008,Shibata_2015,Hu_2022,Chandramouli_2025,Gupta_2025,Yang_2026}. At present, most studies of lensed GW waveforms rely on one of two approaches: the geometric-optics approximation or the Kirchhoff diffraction integral \cite{Lin_2023,Piorkowska_2013,Yuan_2026,Shan_2025,Shan_2025_b}. In the geometric-optics limit, the lensing effect is described by a set of discrete images, each characterized by a frequency-independent magnification factor and phase shift \cite{Isaacson_1968,Keeton_2005,Sereno_2006}. Although this approximation is valid when the GW wavelength is much smaller than the characteristic curvature scale of the lens, it suffers from several notable limitations. First, it predicts strictly frequency-independent amplification factors, which can lead to degeneracies between lens and intrinsic source parameters \cite{Ezquiaga_2021}. Second, it cannot capture inherently wave-effect phenomena such as diffraction and interference. Third, and perhaps most critically for the present work, it completely neglects the tensorial nature and polarization structure of GWs. The Kirchhoff diffraction integral, by contrast, offers a wave-optics description and has been widely employed to model lensing in the long-wavelength regime \cite{Baraldo_1999,Takahashi_2003,Guo_2020}. Nevertheless, this method is not derived from first principles in general relativity; instead, it treats GWs as scalar waves propagating on a fixed background spacetime, thereby neglecting their spin-2 nature \cite{Cusin_2020,Harte_2019a,Harte_2019b}. Moreover, in practical implementations, evaluating the diffraction integral itself relies on geometric-optics approximations, which further limit the approach's consistency and accuracy. In conclusion, the above shortcomings strongly motivate the development of more accurate theoretical models for lensed GW signals.

A first-principle treatment of GW lensing would require solving the full problem of GW propagation in curved spacetime, for instance, the scattering of metric perturbations off a black hole background \cite{Dolan_2008a,Chan_2025,Pijnenburg_2024b,Saketh_2025,Li_2025_b,Li_2025c}. This is a notoriously difficult problem. Analytical approaches to black hole scattering are typically restricted to highly symmetric spacetimes and limited frequency regimes \cite{Li_2025_b,Li_2025c}. Simultaneously, the general-relativistic simulations used by Refs.\,\cite{He_2022,Yin_2024}, though powerful in principle, remain computationally expensive and are challenging to extend to realistic astrophysical configurations or to a wide range of lens models.

This work pursues an alternative strategy: extending the geometric-optics approximation beyond leading order to systematically incorporate higher-order corrections \cite{Isaacson_1968}. In this context, the work of Cusin and collaborators provides a systematic perturbative framework based on a WKB-type expansion of the GW field \cite{Cusin_2020,Dalang_2022,Garoffolo_2024,Menadeo_2025,Postiglione_2025}. In this formalism, the GW metric is decomposed into a rapidly-varying phase and a slowly-varying amplitude tensor, and the GW equations are solved order by order along null geodesics. At leading order, one recovers the standard geometric-optics limit, where GWs propagate along null geodesics, and the polarization tensors are purely transverse. At subleading order, the background curvature sources corrections to the amplitude, leading to qualitatively new physical effects. In particular, diffraction induces deviations from the perfectly transverse nature of the polarization tensor, effectively ``smearing'' the polarization plane and generating apparent vector and scalar polarization components \cite{Cusin_2020,Dalang_2022,Garoffolo_2024,Menadeo_2025,Postiglione_2025}. Subsequent studies have shown that, in simple lensing configurations, such polarization distortions are controlled by a small parameter of order $M\lambdabbar/L^2$ (with $M$ being the lens mass, $\lambdabbar$ the GW wavelength, and $L$ the angular momentum of gravitons), and thus remain subdominant but potentially observable in suitable regimes \cite{Dalang_2022}. This framework, therefore, provides a first-principles-based description of GW propagation, bridging the gap between geometric optics and full-wave treatments while consistently incorporating the spin-2 nature of GWs.

The main objective of this work is to refine the above theoretical description by systematically improving the perturbative framework developed by Cusin and collaborators \cite{Cusin_2020,Dalang_2022}. In particular, we adopt the full Newman–Penrose (NP) formalism \cite{Newman_Penrose_1962,Chandrasekhar1983}, which offers a natural and gauge-invariant language for describing the propagation of spin-2 fields in curved spacetime \cite{Isaacson_1968}. By decomposing the GW metric in terms of an NP tetrad, we derive the evolution equations of the GW amplitude along null geodesics and the explicit expressions of the linearized Weyl tensor. However, the solution of the evolution equations requires the calculation of a set of geometric quantities, e.g., the spin coefficients and their gradients. To this end, we propose the method of solving the geodesic deviation (GD) equation to calculate the spin coefficients \cite{Pineault_1977,Frittelli_2000a,Frittelli_2000b,Gallo_2011,Dolan_2018b,Boero_2019}, and follow the approach of Dolan to derive the transport equations for the gradients of the spin coefficients and the GW amplitudes  \cite{Dolan_2018a,Dolan_2018b,Shipley_2019,Bruyere_2026}. These auxiliary equations serve to close the entire solvable system. The above theoretical framework can, in principle, be applied to any background spacetime that is suitable for the geometric-optics expansion. We then apply this refined theoretical framework to Schwarzschild lensing, demonstrating the evolution of GW amplitudes and Weyl scalars along the trajectories of gravitons. This work also provides a supplementary discussion of the caustic points and the regularization procedure used to numerically eliminate the singularities. Based on the rigorous theoretical framework and numerical solutions, we verify that the GW-background coupling at subleading order does indeed smear the polarization plane, leading to the emergence of non-transverse apparent vector and scalar polarization modes \cite{Cusin_2020,Dalang_2022,Garoffolo_2024,Menadeo_2025,Postiglione_2025}.

This paper is organized as follows. In Sec.\,\ref{sec:propagation-equations}, we review the GW propagation equations on the curved backgrounds, together with the gauge conditions, the eikonal expansion, GW polarization, and their corresponding NP formulation. In Sec.\,\ref{sec:auxiliary}, we derive the three groups of auxiliary equations to ensure the system is solvable. Sec.\,\ref{sec:lensing} applies the well-established framework to Schwarzschild lensing. The corresponding numerical results, including the evolution of GW polarizations, are presented in Sec.\,\ref{sec:numerical}. A brief discussion is presented in Sec.\,\ref{sec:conclusion}. Throughout the paper, we work in geometric units in which $c=G=1$, where $c$ is the speed of light in vacuum and $G$ is the gravitational constant.

\section{\label{sec:propagation-equations}Propagation Equations}

\subsection{\label{subsec:linearized-Einstein-equation}Linearized Einstein equation}

This work focuses on linear gravitational perturbations, separating the spacetime into a background $\gamma_{\mu\nu}$ and a GW perturbation $h_{\mu\nu}$, i.e., $g_{\mu\nu}=\gamma_{\mu\nu}+h_{\mu\nu}$. The amplitude of the GW perturbation is assumed to be much smaller than that of the background curvature. In terms of the trace-reversed tensor, defined by
\begin{equation}
\psi_{\mu\nu}=h_{\mu\nu}-\frac{1}{2}\gamma_{\mu\nu}h,
\end{equation}
the Einstein equation reads \cite{MTW,Wald_1984,Isaacson_1968}
\begin{equation}
\label{propagation-equation-GW-curved-background-2}
\Box^2\psi_{\mu\nu}+2R^{(0)}_{\alpha\nu\beta\mu}\psi^{\alpha\beta}=0.
\end{equation}
in vacuum, where the perturbation tensor is required to satisfy the Lorenz and traceless gauges, i.e.,
\begin{equation}
\label{gauge}
\nabla_{\mu}\psi^{\mu\alpha}=\gamma^{\mu\nu}\psi_{\mu\nu}=0,
\end{equation}
$\nabla_{\mu}$ is the covariant derivative compatible with the background spacetime $\gamma_{\mu\nu}$, $\Box^2\equiv\nabla^{\alpha}\nabla_{\alpha}$ is the d'Alembertian, and $R^{(0)}_{\alpha\nu\beta\mu}$ is the Riemann tensor of the background.
	
\subsection{Geometric-optics expansion}

In typical strong-lensing scenarios, the GW wavelength is much smaller than the background curvature radius. Denoting the typical wavelength of the GW by $\lambdabbar$ and the characteristic curvature scale of the background by $\mathcal{R}$, the short-wavelength condition is expressed as $\lambdabbar\ll\mathcal{R}$. Under this approximation, we formally expand the GW metric as \cite{MTW,Hou_2019,Hou_2020b,Cusin_2020,Dalang_2022}
\begin{equation}
\label{eikonal-expansion}
\psi_{\mu\nu}=\mathrm{Re}\,\Big\{\left[A_{\mu\nu}+\epsilon B_{\mu\nu}+\cdots\right]e^{i\Phi/\epsilon}\Big\}.
\end{equation}
This expansion is usually referred to as the short-wave or eikonal expansion. In Eq.\,(\ref{eikonal-expansion}), $\epsilon\sim\lambdabbar/\mathcal{R}$ is a bookkeeping parameter. The scalar field $\Phi$ represents the GW phase, related to the 4-wave vector by
\begin{equation}
\label{definition-wave vector}
k_{\mu}=-\nabla_{\mu}\Phi,
\end{equation} 
Inserting expansion (\ref{eikonal-expansion}) into Eq.\,(\ref{propagation-equation-GW-curved-background-2}), one obtains at the lowest order ($\epsilon^{-2}$) the null condition $k^{\alpha}k_{\alpha}=0$, which shows that GWs propagate at the speed of light. One can further show that the wave vector satisfies the geodesic equation $k^{\alpha}\nabla_{\alpha}k_{\mu}=0$. The $\epsilon^{-1}$ order gives the evolution of the leading-order (LO) amplitude $A_{\mu\nu}$, satisfying \cite{MTW}
\begin{equation}
\label{evolution-equation-A}
k_{\alpha}\nabla^{\alpha}A_{\mu\nu}+\frac{1}{2}(\nabla_{\alpha}k^{\alpha})A_{\mu\nu}=0.
\end{equation}
Furthermore, at $\epsilon^{0}$ order, the evolution of the subleading-order (SLO) amplitude $B_{\mu\nu}$ is given by
\begin{equation}
\label{evolution-equation-B}
k_{\alpha}\nabla^{\alpha}B_{\mu\nu}+\frac{1}{2}(\nabla_{\alpha}k^{\alpha})B_{\mu\nu}
=-\frac{i}{2}\left[\Box^2A_{\mu\nu}+2R^{(0)}_{\mu\alpha\nu\beta}A^{\alpha\beta}\right].
\end{equation}

Under the short-wavelength expansion, the Lorenz and traceless gauges (\ref{gauge}) are written as
\begin{equation}
\label{Lorenz-gauge}
k^{\mu}A_{\mu\nu}=0,\quad\text{and}\quad
k^{\mu}B_{\mu\nu}+i\nabla^{\mu}A_{\mu\nu}=0,
\end{equation}
and
\begin{equation}
\label{traceless-gauge}
\gamma^{\mu\nu}A_{\mu\nu}
=\gamma^{\mu\nu}B_{\mu\nu}=0.
\end{equation}
Eq.\,(\ref{Lorenz-gauge}) shows that the LO amplitude is orthogonal to the wave vector, implying that GWs propagate as transverse waves at this order. This conclusion, however, does not extend to the SLO amplitude: the non-orthogonality between the wave vector and the SLO amplitude indicates that GWs acquire non-transverse oscillation components at subleading order.

\subsection{Newman-Penrose decomposition}
To solve Eqs.\,(\ref{evolution-equation-A},\,\ref{evolution-equation-B}) and extract the polarization characteristics of GWs more conveniently, we project the GW amplitudes and propagation equations onto an NP tetrad \cite{Hou_2019,Hou_2020b,Cusin_2020,Dalang_2022}. This tetrad consists of four null legs \cite{Chandrasekhar1983}: 
\begin{equation}
e_{(a)}^{\mu}=\left\{k^{\mu},n^{\mu},m^{\mu},\bar{m}^{\mu}\right\}.
\end{equation}
In this work, to investigate GW propagation, the null tetrad is required to be parallel-transported along the null geodesics, i.e.,
\begin{equation}
k^{\alpha}\nabla_{\alpha}e_{(a)}^{\mu}=0.   
\end{equation}
Greek indices are spacetime components, and bracketed Latin indices are tetrad components. In this convention, spacetime indices of any physical quantities are raised and lowered with $\gamma_{\mu\nu}$. The tetrad indices are raised and lowered with $\eta_{(a)(b)}$, as defined in Eq.\,(\ref{orthogonality}). The first leg is identified with the GW wave vector, while the second leg points in the spatially opposite direction. The third and fourth legs span the polarization plane and are complex conjugates of each other. The null condition is $-\bm{k}\cdot\bm{n}=\bm{m}\cdot\bar{\bm{m}}=1$, and all other inner products vanish. Associated with these four null directions, the directional-derivative operators are defined by
\begin{equation}
\label{directional-derivative}
\bm{\mathcal{D}}\equiv k^{\mu}\partial_{\mu},\quad
\bm{\Delta}\equiv n^{\mu}\partial_{\mu},\quad
\bm{\delta}\equiv m^{\mu}\partial_{\mu},\quad
\bar{\bm{\delta}}\equiv \bar{m}^{\mu}\partial_{\mu},
\end{equation}
Note that the operator $\bm{\mathcal{D}}$ is equivalent to the derivative with respect to an affine parameter $s$, i.e., $\bm{\mathcal{D}}=\dd/\dd{s}$.

In the main text, we briefly introduce the notation of the NP formalism, and more details are provided in Appendix \ref{app:NP-formalism}. The spacetime geometry is characterized by a set of NP scalars: the spin coefficients (see Eq.\,(\ref{spin-coefficient-definition}))
$$\rho,\,\sigma,\,\alpha,\,\beta,\,\lambda,\,\mu,\,\gamma,\,\nu,\,\tau,\,\kappa,\,\varepsilon,\,\pi,$$
the Weyl scalars $\Psi_n$ $(n=0,1,2,3,4)$ (see Eq.\,(\ref{Weyl-scalars-definition})), and the Ricci scalars $\Phi_{nm}$ $(n,m=0,1,2)$, which vanish because we focus on a vacuum background. Among these, $\kappa=\varepsilon=\pi=0$ and $\tau=\bar{\alpha}+\beta$ as a consequence of the parallel-transport condition. In addition, $\rho$ is real as a consequence of the scalar nature of the phase $\Phi$. The dynamics of spacetime is governed by the NP equations, the commutators of the directional derivatives (see Eq.\,(\ref{commutators})), and the Bianchi identities (see Eq.\,(\ref{Bianchi-ID})). In particular, the NP equations reduce to the Sachs equations for the parallel-transported tetrad (see Eq.\,(\ref{Sachs-equations})).

Following Ref.\,\cite{Cusin_2020}, we define the basis tensors as 
\begin{equation}
\label{basis-tensor-definition}
\Big[\Theta^{(a)(b)}\Big]_{\mu\nu}\equiv\frac{1}{2}\left[e^{(a)}_{\mu}e^{(b)}_{\nu}+e^{(a)}_{\nu}e^{(b)}_{\mu}\right],
\end{equation}
which is symmetric in both the spacetime indices $\mu,\nu$ and the tetrad indices $(a),(b)$. The orthogonality relation for the basis tensors reads
\begin{equation}
\label{basis-tensor-orthogonality}
\Big[\Theta_{(a)(b)}\Big]_{\mu\nu}
\Big[\Theta^{(c)(d)}\Big]^{\mu\nu}
=\frac{1}{2}
\left[\delta_{(a)}^{(c)}
\delta_{(b)}^{(d)}
+\delta_{(a)}^{(d)}
\delta_{(b)}^{(c)}\right].
\end{equation}
An arbitrary symmetric tensor, e.g., $A_{\mu\nu}$, can be decomposed in this basis as
\begin{equation}
\label{amplitude-A-decomposition}
A_{\mu\nu}=A_{(a)(b)}\left[\Theta^{(a)(b)}\right]_{\mu\nu},
\end{equation}
with
\begin{equation}
A_{(a)(b)}=A_{\mu\nu}\left[\Theta_{(a)(b)}\right]^{\mu\nu}.
\end{equation}
Using the formalism described above, we simplify the gauge conditions (see Sec.\,\ref{subsec:gauge}), the expressions of GW polarizations (see Sec.\,\ref{subsec:polarization}), and the equations of GW propagation (see Sec.\,\ref{subsec:equation}) in the following subsections.

\subsection{\label{subsec:gauge}Gauge conditions}
In subsection \ref{subsec:linearized-Einstein-equation}, we have applied the Lorenz and traceless gauges to the linearized Einstein equation, thereby eliminating five of the eight residual degrees of freedom (DoFs). In terms of the NP tetrad, these gauge conditions follow from Eqs.\,(\ref{Lorenz-gauge},\,\ref{traceless-gauge}) and reduce to
\begin{equation}
A_{kk}=A_{kn}=A_{km}=A_{k\bar{m}}=A_{m\bar{m}}=0,
\end{equation}
from Eqs.\,(\ref{Lorenz-gauge},\,\ref{traceless-gauge}). Three residual DoFs remain, corresponding to three undetermined components in the gauge vector. The gauge transformation is written as $h_{\mu\nu}\rightarrow h_{\mu\nu}-(\nabla_{\mu}\Sigma_{\nu}+\nabla_{\nu}\Sigma_{\mu})$. As a consequence of the Lorenz gauge, the gauge vector $\Sigma_{\mu}$ is required to obey the wave equation $\Box^2\Sigma_{\mu}=0$. Under the geometric-optics expansion, the high-frequency solution for $\Sigma_{\mu}$ takes the form
\begin{equation}
\label{eikonal-expansion-gauge-vector}
\Sigma_{\mu}=\mathrm{Re}\left[\epsilon w_{\mu}e^{i\Phi/\epsilon}+\mathcal{O}(\epsilon)\right].
\end{equation}
Combining this with the expansion for $\psi_{\mu\nu}$ in Eq.\,(\ref{eikonal-expansion}), the transformation rule for the LO amplitude is
\begin{equation}
\label{transformation-rule-A}
A_{\mu\nu}\rightarrow \tilde{A}_{\mu\nu}=A_{\mu\nu}
-i\left(k_{\mu}w_{\nu}+k_{\nu}w_{\mu}\right).
\end{equation}
To fix gauge and set three components of the transformed amplitude to zero, we project the transformation rule (\ref{transformation-rule-A}) onto the NP tetrad, obtaining
\begin{equation}
\tilde{A}_{(a)(b)}=A_{(a)(b)}
-i\left[k_{(a)}w_{(b)}+k_{(b)}w_{(a)}\right].
\end{equation}
Under the lowest-order expansion, the covariant derivative operator $\nabla_{\mu}$ reduces to $-ik_{\mu}$. Under NP projection, the tetrad components of the wave vector are given by $k_{n}=-1$ and $k_{k}=k_{m}=k_{\bar{m}}=0$. One thus obtains
\begin{equation}
\begin{aligned}
\tilde{A}_{nn}&=A_{nn}+2iw_{n},\\
\tilde{A}_{nm}&=A_{nm}+iw_{m},\\
\tilde{A}_{n\bar{m}}&=A_{n\bar{m}}+iw_{\bar{m}}.
\end{aligned}
\end{equation}
Requiring the above three equations to vanish uniquely determines the gauge vector. Additionally, the gauge vector is constrained by the traceless condition $k^{\alpha}w_{\alpha}=w_{k}=0$. In conclusion, the gauge vector is adopted as following,
\begin{equation}
w_{(a)}=i\left\{0,\frac{1}{2}A_{nn},A_{nm},A_{n\bar{m}}\right\},
\end{equation}
at the leading order, which ensures $\tilde{A}_{nn}=\tilde{A}_{nm}=\tilde{A}_{n\bar{m}}=0$ \cite{Cusin_2020}. These three gauge conditions remove the trace of the metric perturbation and all longitudinal components (i.e., those spatially parallel or antiparallel to the wave vector), which is equivalent to imposing the transverse-traceless gauge on the GW metrics. At this stage, the gauge fixing is complete, and the full set of gauge conditions for the LO amplitude can be summarized as
\begin{equation}
\label{gauge-A}
A_{kk}=A_{kn}=A_{km}=A_{k\bar{m}}=A_{nn}=A_{nm}=A_{n\bar{m}}=A_{m\bar{m}}=0.
\end{equation}

The gauge conditions (\ref{Lorenz-gauge},\,\ref{traceless-gauge}) for the SLO amplitude can also be expressed in NP formalism. The Lorenz gauge (\ref{Lorenz-gauge}) is written as
\begin{subequations}
\label{gauge_B_1}
\begin{align}
&B_{kk}=i(\bar{\sigma}A_{mm}+\sigma A_{\bar{m}\bar{m}}),\\
&B_{kn}
=-i(\lambda A_{mm}+\bar{\lambda}A_{\bar{m}\bar{m}}),\\
&B_{km}=-i\left[\bar{\bm{\delta}}+2(\alpha-\bar{\beta})+\bar{\tau}\right]A_{mm},\\
&B_{k\bar{m}}=-i[\bm{\delta}+2(\bar{\alpha}-\beta)+\tau]A_{\bar{m}\bar{m}}.
\end{align}
\end{subequations}
The traceless gauge (\ref{traceless-gauge}) becomes
\begin{equation}
\label{gauge_B_2}
B_{m\bar{m}}=B_{kn}.
\end{equation}
These two sets of relations are used to simplify the expressions of GW polarization and the equations of GW evolution in the following sections.

\subsection{\label{subsec:polarization}Gravitational wave polarizations}
We now turn our attention to the GW polarizations in the NP formalism. On the flat background, one usually uses the Riemann tensor to express GW polarization because of its gauge invariance \cite{Will_2014}. However, this is no longer valid on the curved backgrounds \cite{Isaacson_1968}. We first prove that the gauge invariance of the Riemann tensor remains valid up to the SLO geometric optics expansion, and then present the explicit expressions of the Riemann tensor in terms of NP scalars.

The rank-$(0,4)$ Riemann tensor of linear GWs is \cite{Cusin_2020}
\begin{equation}
\label{Riemann-tensor-GW}
R^{(1)}_{\alpha\beta\mu\nu}
=2\nabla_{[\mu|}
\nabla_{[\beta}\psi_{\alpha]|\nu]}
+\psi^{\gamma}_{[\alpha|}
R^{(0)}_{\gamma|\beta]\mu\nu},
\end{equation}
where the superscript $^{(1)}$ denotes the first-order perturbation (GW). Under the gauge transformation $h_{\mu\nu}\rightarrow h_{\mu\nu}-(\nabla_{\mu}\Sigma_{\nu}+\nabla_{\nu}\Sigma_{\mu})$, the linearized Riemann tensor transforms as \cite{Isaacson_1968}
\begin{equation}
\label{Riemann-tensor-GW-gauge-transformation}
R^{(1)}_{\alpha\beta\mu\nu}
\rightarrow R^{(1)}_{\alpha\beta\mu\nu}
-\Big[\left(\nabla_{\gamma}R^{(0)}_{\alpha\beta\mu\nu}\right)
\Sigma^{\gamma}
+\left(\nabla_{\alpha}\Sigma^{\gamma}\right)R^{(0)}_{\gamma\beta\mu\nu}
+\left(\nabla_{\beta}\Sigma^{\gamma}\right)R^{(0)}_{\alpha\gamma\mu\nu}
+\left(\nabla_{\mu}\Sigma^{\gamma}\right)R^{(0)}_{\alpha\beta\gamma\nu}
+\left(\nabla_{\nu}\Sigma^{\gamma}\right)R^{(0)}_{\alpha\beta\mu\gamma}\Big].
\end{equation}
To assess the orders of magnitude of the polarization modes and the gauge transformation, we proceed as follows. On the one hand, from Eq.\,(\ref{Riemann-tensor-GW}), we find that the first term contains contributions of order $\mathcal{O}(h/\lambdabar^2)$, $\mathcal{O}(h/\lambdabar\mathcal{R})$, and $\mathcal{O}(h/\mathcal{R}^2)$, while the second term is only $\sim\mathcal{O}(h/\mathcal{R}^2)$. Up to subleading order, only the terms $h/\lambdabar^2$ and $h/\lambdabar\mathcal{R}$ should be considered. Therefore, the second term does not contribute to the final expression of the polarizations at the considered orders. On the other hand, from Eq.\,(\ref{Riemann-tensor-GW-gauge-transformation}), the lowest-order term in the squared brackets is of order $\mathcal{O}(h/\mathcal{R}^2)$, which likewise does not affect the SLO polarizations. In conclusion, the Riemann tensor provides a gauge-invariant characterization of the polarization modes at the considered orders of this work.

Substituting the expansion (\ref{eikonal-expansion}) into the linearized Riemann tensor (\ref{Riemann-tensor-GW}), we obtain \cite{Cusin_2020}
\begin{equation}
R_{\alpha\beta\mu\nu}^{(1)}
=\epsilon^{-2}R_{\alpha\beta\mu\nu}^{(1),{\rm LO}}+\epsilon^{-1}R_{\alpha\beta\mu\nu}^{(1),{\rm SLO}},
\end{equation}
\begin{equation}
R_{\alpha\beta\mu\nu}^{(1),{\rm LO}}=\mathrm{Re}\,\Big\{-2k_{[\mu|}k_{[\beta}A_{\alpha]|\nu]}e^{i\Phi/\epsilon}\Big\},
\end{equation}
\begin{equation}
R_{\alpha\beta\mu\nu}^{(1),{\rm SLO}}
=\mathrm{Re}\,\Bigg\{\Big[-2i\nabla_{[\beta}A_{\alpha][\nu}k_{\mu]}
-2i\nabla_{[\mu}A_{\nu][\alpha}k_{\beta]}
+2i(\nabla_{[\mu|}k_{[\beta})A_{\alpha]|\nu]}
-2k_{[\mu|}k_{[\beta}B_{\alpha]|\nu]}\Big]e^{i\Phi/\epsilon}\Bigg\},
\end{equation}
where the anti-symmetrization brackets are defined as $t_{[\alpha\beta]}\equiv(1/2)(t_{\alpha\beta}-t_{\beta\alpha})$ and $t_{[\mu|[\alpha\beta]|\nu]}\equiv(1/2)(t_{\mu[\alpha\beta]\nu}-t_{\nu[\alpha\beta]\mu})$.

The Weyl scalars, defined in Eq.\,(\ref{Weyl-scalars-definition}), are obtained from the NP projection of the linearized Riemann tensor,
\begin{equation}
\label{Riemann-tensor-GW-project}
R^{(1)}_{(a)(b)(c)(d)}
=e_{(a)}^\alpha
e_{(b)}^\beta
e_{(c)}^\mu
e_{(d)}^\nu
R^{(1)}_{\alpha\beta\mu\nu}.
\end{equation}
After a straightforward but lengthy calculation, we obtain
\begin{equation}
\Psi_{n}^{(1)}
=\epsilon^{-2}\Psi_{n,{\rm LO}}^{(1)}
+\epsilon^{-1}\Psi_{n,{\rm SLO}}^{(1)},
\end{equation}
where
\begin{equation}
\label{Psi-LO}
\Psi_{0,{\rm LO}}^{(1)}
=\Psi_{1,{\rm LO}}^{(1)}
=\Psi_{2,{\rm LO}}^{(1)}
=\Psi_{3,{\rm LO}}^{(1)}=0,
\quad\text{and}\quad
\Psi_{4,\mathrm{LO}}^{(1)}=-\frac{1}{2}
\Big\{\bar{m}^{\mu}\bar{m}^{\nu}\,\mathrm{Re}(A_{\mu\nu}e^{i\Phi/\epsilon})\Big\},
\end{equation}
at the LO approximation. At subleading order, the remaining Weyl scalars are given by
\begin{equation}
\label{Psi-0-1-SLO}
\Psi_{0,{\rm SLO}}^{(1)}=\Psi_{1,{\rm SLO}}^{(1)}=0,
\end{equation}
\begin{equation}
\label{Psi-2-SLO}
\Psi_{2,\mathrm{SLO}}^{(1)}=-\frac{i\sigma}{2}\Big\{\bar{m}^{\mu}\bar{m}^{\nu}\,\mathrm{Im}(A_{\mu\nu}e^{i\Phi/\epsilon})\Big\},
\end{equation}
\begin{equation}
\label{Psi-3-SLO}
\Psi_{3,\mathrm{SLO}}^{(1)}=\frac{i}{2}(\bar{\alpha}-3\beta+\bm{\delta})
\Big\{\bar{m}^{\mu}\bar{m}^{\nu}\,\mathrm{Im}(A_{\mu\nu}e^{i\Phi/\epsilon})\Big\},
\end{equation}
\begin{equation}
\label{Psi-4-SLO}
\Psi_{4,\mathrm{SLO}}^{(1)}
=-\frac{1}{2}\Big\{\bar{m}^{\mu}\bar{m}^{\nu}\,\mathrm{Re}(B_{\mu\nu}e^{i\Phi/\epsilon})\Big\}
+i\left[\bm{\Delta}-\left(\frac{5}{2}\gamma-\frac{3}{2}\bar{\gamma}+\bar{\mu}\right)\right]
\Big\{\bar{m}^{\mu}\bar{m}^{\nu}\,\mathrm{Im}(A_{\mu\nu}e^{i\Phi/\epsilon})\Big\}.
\end{equation}
Note that, under the vacuum assumption, all Ricci scalars vanish. Our result in Eq.\,(\ref{Psi-2-SLO}) is consistent with the result of Ref.\,\cite{Harte_2019a}, and Eq.\,(\ref{Psi-4-SLO}) is consistent with Ref.\,\cite{Hou_2020b}. When deriving Eqs.\,(\ref{Psi-0-1-SLO},\,\ref{Psi-2-SLO},\,\ref{Psi-3-SLO},\,\ref{Psi-4-SLO}), we have used the gauge conditions (\ref{gauge_B_1},\,\ref{gauge_B_2}).

As shown in Eq.\,(\ref{Psi-LO}), only one non-vanishing Weyl scalar appears at leading order, representing the right-handed polarization of the GWs (the left-handed polarization is represented by the complex conjugate of $\Psi^{(1)}_{4,\mathrm{LO}}$), which shows a similar structure to a plane wave on the flat backgrounds. This is because, at this order, the polarization plane remains transverse to the propagation direction, and the transverse-traceless gauge has been applied to the LO amplitude. The SLO amplitudes $B_{mm}$ and $B_{\bar{m}\bar{m}}$ appear only in $\Psi_{4,\text{SLO}}^{(1)}$. Thus, only these two components need to be solved from Eq.\,(\ref{evolution-equation-B}); they represent the gauge-invariant sector of the SLO amplitude tensor.

The most important observation regarding GW polarization is that the Weyl scalars $\Psi_2^{(1)}$ and $\Psi_3^{(1)}$, which represent scalar and vector polarization modes respectively, do not vanish. Meanwhile, it can be shown that it is impossible to eliminate all of them simultaneously via a tetrad transformation \cite{Cusin_2020,Carmeli_1976}. This originates from the coupling between GWs and the curved spacetime background. Before the GW signals approach the lensing object, the wavefront is typically spherical, and once the GWs have propagated to a location sufficiently distant from the source, the wavefront can be safely approximated as planar. This ensures that $B_{\mu\nu}$ and $k^{\mu}$ are strictly orthogonal, thereby avoiding non-zero $\Psi_2^{(1)}$ and $\Psi_3^{(1)}$. During the lensing process, as calculated by the black hole scattering method, the external gravitational field significantly distorts the wavefront \cite{Li_2025_b,Li_2025c}. The non-orthogonality between $B_{\mu\nu}$ and $k^{\mu}$, i.e., $k^{\mu}B_{\mu\nu}\neq0$, induces non-zero $\Psi_2^{(1)}$ and $\Psi_3^{(1)}$, as demonstrated by our derivation. However, it should be emphasized that the polarization modes represented by these Weyl scalars are apparent and do not correspond to any new physical degrees of freedom: mathematically, $\Psi_2^{(1)}$ and $\Psi_3^{(1)}$ are fully determined by the LO amplitude $A_{\mu\nu}$, and it will be explained that $B_{\mu\nu}$ is ultimately determined by $A_{\mu\nu}$ as well. Apart from the initial $+$ and $\times$ modes, there are no physical quantities that need to be specified by hand. Physically, this study remains within the framework of general relativity, where general covariance and the Bianchi identity guarantee that the intrinsic degrees of freedom of the Einstein equations are exactly two.

\subsection{\label{subsec:equation}Equations of amplitude evolution}

We now derive the evolution equations for the GW amplitudes in the NP formalism. Multiplying both sides of Eqs.\,(\ref{evolution-equation-A},\,\ref{evolution-equation-B}) by the basis tensor (\ref{basis-tensor-definition}), we obtain
\begin{equation}
\label{evolution-equation-A-NP}
\bm{\mathcal{D}}A_{(a)(b)}+\rho A_{(a)(b)}=0,
\end{equation}
and
\begin{equation}
\label{evolution-equation-B-NP}
\bm{\mathcal{D}}B_{(a)(b)}+\rho B_{(a)(b)}=\mathcal{S}_{(a)(b)}.
\end{equation}
The source terms in Eq.\,(\ref{evolution-equation-B-NP}) are
\begin{subequations}
\begin{align}
&\begin{aligned}
2i\mathcal{S}_{mm}
&=(\bm{\delta}\bar{\bm{\delta}}
+\bar{\bm{\delta}}\bm{\delta})A_{mm}
+\left[(7\alpha-3\bar{\beta})\bm{\delta}
+(-\bar{\alpha}+5\beta)\bar{\bm{\delta}}\right]A_{mm}\\
&+\left[2(\bm{\delta}\alpha)-2(\bar{\bm{\delta}}\beta)^*+4(\bar{\bm{\delta}}\beta)
-4\alpha\bar{\alpha}+20\alpha\beta-8\beta\bar{\beta}-\rho(3\mu+\bar{\mu})-2\bar{\lambda}\bar{\sigma}-6\Psi^{(0)}_2\right]A_{mm},
\end{aligned}\\
&\begin{aligned}
2i\mathcal{S}_{\bar{m}\bar{m}}
&=(\bm{\delta}\bar{\bm{\delta}}
+\bar{\bm{\delta}}\bm{\delta})A_{\bar{m}\bar{m}}
+\left[(7\bar{\alpha}-3\beta)\bar{\bm{\delta}}+(-\alpha+5\bar{\beta})\bm{\delta}\right]A_{\bar{m}\bar{m}}\\
&+\left[-2(\bm{\delta}\alpha)+2(\bar{\bm{\delta}}\beta)^*+4(\bm{\delta}\alpha)^*
-4\alpha\bar{\alpha}+4\alpha\beta+16\bar{\alpha}\bar{\beta}-8\beta\bar{\beta}-\rho(3\mu+\bar{\mu})-2\bar{\lambda}\bar{\sigma}-2\Psi^{(0)}_2-4\bar{\Psi}^{(0)}_2\right]A_{\bar{m}\bar{m}}.
\end{aligned}
\end{align}
\end{subequations}
Here and throughout, the superscript $^*$ denotes complex conjugation. As shown above, other components of $B_{(a)(b)}$, such as $B_{nm}$, are unnecessary to calculate the GW Weyl scalars. The LO evolution equation (\ref{evolution-equation-A-NP}) is consistent with the results of \,\cite{Dolan_2018a,Dolan_2018b,Shipley_2019}. The SLO equations (\ref{evolution-equation-B-NP}) are similar to those given in Ref.\,\cite{Cusin_2020}. In this work, we recast them entirely in terms of NP scalars, so that the evolution equations for different metric components are fully decoupled.

\section{\label{sec:auxiliary}Auxiliary system of equations}
Before solving for the GW amplitudes, one has to compute several sets of geometric quantities, namely:

(1) the spin coefficients, $\rho$, $\sigma$, $\alpha$, $\beta$, $\cdots$;

(2) the gradients of the spin coefficients, $\bm{\delta}\rho$, $\bm{\delta}\bm{\delta}\rho$, $\cdots$;

(3) the gradients of the GW amplitudes, $\bm{\delta}A_{mm}$, $\bm{\delta}\bm{\delta}A_{mm}$, $\cdots$.

Although Eqs.\,(\ref{spin-coefficient-definition}) and (\ref{directional-derivative}) provide clear definitions of the spin coefficients and the directional derivatives, respectively, evaluating them in practice is far from straightforward. For example, the explicit definition of $\rho$ is $\rho\equiv\gamma_{(3)(1)(4)}=m^{\mu}(\nabla_{\mu}k_{\nu})\bar{m}^{\nu}$, which represents the transverse derivative of the 4-wave vector. However, $k_{\mu}$, the solution to null geodesic equations, depends not only on the spacetime coordinates but also on a set of constants of motion, such as the angular momentum and the Carter constant. These quantities are conserved along the given geodesic, but vary across neighboring geodesics and must therefore be treated as variables when calculating transverse derivatives. However, expressing the constants of motion explicitly as functions of the spacetime coordinates is rather difficult. An indirect approach is therefore required to calculate the spin coefficients and their gradients. In this work, together with solving the expansion and shear of null congruences, we use the Sachs equations (\ref{Sachs-equations}) to solve for the spin coefficients and employ the transport equations to obtain their gradients \cite{Dolan_2018a,Dolan_2018b,Shipley_2019}. Another groups of transport equations are also introduced to calculate the gradients of the GW amplitudes.

\subsection{Computing spin coefficients through null congruence}

The Sachs equations describe the evolution of null geodesic congruences \cite{Frittelli_2000a,Frittelli_2000b,Gallo_2011,Dolan_2018b,Boero_2019,Li_2025}. Conversely, the theory of null congruence provides a systematic method for constructing solutions to the Sachs equations. We denote the tetrad legs as functions of the affine parameter $s$ and a set of constants of motion $\mu^{a}$, i.e., $\bm{e}_{(a)}=\bm{e}_{(a)}(s,\mu^{a})$. For simplicity, we consider two neighboring GW rays that are sufficiently close together, with trajectories denoted by
\begin{equation}
x^{\alpha}(s;\mu^a),\quad\text{and}\quad
x^{\alpha}(s+\delta s;\mu^a+\delta\mu^a),
\end{equation}
where $\delta s$ and $\delta\mu^a$ represent small changes in the affine parameter and the constants of motion for the two graviton trajectories. The GD vector is defined as the coordinate difference between these two gravitons. Expanding to the first order in the small parameters $\delta s$ and $\delta\mu^a$ yields
\begin{equation}
\xi^{\alpha}
\equiv x^{\alpha}(s+\delta s;\mu^a+\delta\mu^a)
-x^{\alpha}(s;\mu^a)
=\frac{\partial x^{\alpha}}{\partial s}(\delta s)
+\sum_a\frac{\partial x^{\alpha}}{\partial\mu^a}(\delta\mu^a)
+\mathcal{O}(\delta s,\delta\mu^a).
\end{equation}

On the one hand, the difference $\delta k^{\nu}$ between the wave vectors associated with $x^{\alpha}$ and $x^{\alpha}+\xi^{\alpha}$ can be expressed in terms of the deviation vector as
\begin{equation}
\label{delta-k-1}
\delta k^{\nu}=(\partial_{\alpha}k^{\nu})\xi^{\alpha}+\mathcal{O}(\xi^{\alpha})
\end{equation}
On the other hand, the same difference can also be expanded directly in $\delta s$ and $\delta\mu^a$, giving
\begin{equation}
\label{delta-k-2}
\delta k^{\nu}
=k^{\alpha}(\partial_{\alpha}\xi^{\nu})
+\mathcal{O}(\xi^{\alpha}),
\end{equation}
Combining Eqs.\,(\ref{delta-k-1}) and (\ref{delta-k-2}) to linear order in $\delta s$ and $\delta\mu^a$ yields the first-order relation
\begin{equation}
\label{GDV-1st-equation-2}
(\nabla_{\alpha}k^{\nu})\xi^\alpha
=k^{\alpha}\nabla_{\alpha}\xi^{\nu},
\end{equation}
Applying the operator $\bm{\mathcal{D}}$ to Eq.\,(\ref{GDV-1st-equation-2}) leads to the GD equation
\begin{equation}
\label{GDV-2nd-equation}
\begin{aligned}
\bm{\mathcal{D}}^2\xi_{\nu}
=-R^{(0)}_{\nu\mu\alpha\beta}k^{\mu}k^{\beta}\xi^{\alpha}.
\end{aligned}
\end{equation}
Equation (\ref{GDV-1st-equation-2}) relates the tensor $\nabla_{\mu}k_{\nu}$ to the deviation vector, while Eq.\,(\ref{GDV-2nd-equation}) governs the evolution of $\xi^{\nu}$ and can be solved directly \cite{Li_2025}. Crucially, this procedure avoids the need to compute derivatives of the wave vector directly.

To solve the GD equation in a convenient form, we decompose $\xi^{\alpha}$ in terms of the NP tetrad as
\begin{equation}
\label{GDVs-NP}
\xi^{\alpha}=\xi^{(a)}e_{(a)}^{\alpha},\quad\text{with}\quad
\xi^{(a)}=\xi^{\alpha}e^{(a)}_{\alpha}.
\end{equation}
Imposing the equal-phase condition $\xi^{n}=k_{\alpha}\xi^{\alpha}=0$ and substituting Eq.\,(\ref{GDVs-NP}) into Eqs.\,(\ref{GDV-1st-equation-2},\,\ref{GDV-2nd-equation}), we obtain
\begin{equation}
\label{GDE-1-NP}
\bm{\mathcal{D}}\xi^{m}
=\rho\,\xi^{m}+\bar{\sigma}\xi^{\bar{m}},
\end{equation}
and
\begin{equation}
\label{GDE-2-NP}
\bm{\mathcal{D}}^2\xi^{m}=\bar{\Psi}^{(0)}_0\xi^{\bar{m}},
\end{equation}
Following our previous work \cite{Li_2025}, by introducing $\xi_{\pm}\equiv\xi^{m}\pm\xi^{\bar{m}}$ and assuming $\Psi_0^{(0)}$ to be real (This is a special case, but it is sufficient for the consideration of this work), we decouple Eq.\,(\ref{GDE-2-NP}) into
\begin{equation}
\label{GDE-2-NP-decouple}
\bm{\mathcal{D}}^2\xi_{\pm}=\pm\Psi^{(0)}_0\xi_{\pm},
\end{equation}
Considering the point-like GW source, we specialize the initial conditions of $\xi_{\pm}$ as $\xi_{\pm}(s=0)=0$ and $\bm{\mathcal{D}}\xi_{\pm}(s=0)=c_{\pm}$. The integration constants $c_{\pm}$ represent the initial opening angle of the null congruence. Re-expressing $\xi^{m}$, $\xi^{\bar{m}}$ in Eq.\,(\ref{GDE-1-NP}) in terms of $\xi^{\pm}$, and demanding the equation hold for arbitrary $c_{\pm}$, the solutions to $\rho$ and $\sigma$ are given by \cite{Li_2025}
\begin{equation}
\label{rho}
\rho=\bm{\mathcal{D}}\ln\sqrt{\left|\xi_{+}\xi_{-}\right|},
\end{equation}
and
\begin{equation}
\label{sigma}
\sigma=\bm{\mathcal{D}}\ln\sqrt{\left|\xi_{+}/\xi_{-}\right|},
\end{equation}
which are independent to the choice of $c_{\pm}$, therefore, we are able to set $c_{\pm}=1$ without loss of generalization. In the physical sense, $\rho$ and $\sigma$ represent the relative change rates of cross-sectional area ($\propto|\xi_{+}\xi_{-}|^{1/2}$) and axis ratio ($\propto|\xi_{+}/\xi_{-}|$) of the null congruences, respectively. With the solutions to $\rho$ and $\sigma$ in hand, the expressions of other spin coefficients are obtained from the Sachs equations (\ref{Sachs-equations}). They are
\begin{equation}
\label{lambda}
\lambda=\frac{1}{2}\left(\frac{\mathcal{Q}_{+}}{\xi_{+}}-\frac{\mathcal{Q}_{-}}{\xi_{-}}\right),
\end{equation}
\begin{equation}
\label{mu}
\mu=\frac{1}{2}\left(\frac{\mathcal{Q}_{+}}{\xi_{+}}+\frac{\mathcal{Q}_{-}}{\xi_{-}}\right),
\end{equation}
\begin{equation}
\label{alpha}
\alpha=\frac{1}{2}\left(\frac{\Pi_{+}}{\xi_{+}}-\frac{\Pi_{-}}{\xi_{-}}\right)
\end{equation}
\begin{equation}
\label{beta}
\beta=\frac{1}{2}\left(\frac{\Pi_{+}}{\xi_{+}}+\frac{\Pi_{-}}{\xi_{-}}\right),
\end{equation}
\begin{equation}
\label{tau}
\tau=\frac{\Pi_{+}}{\xi_{+}}.
\end{equation}
\begin{equation}
\label{gamma}
\gamma=\int_0^{s}\Psi_2^{(0)}\dd{s}
-\int_0^{s}\left(\frac{\Pi_{+}}{\xi_{+}}\right)^2\dd{s}
+c_{\gamma},
\end{equation}
\begin{equation}
\label{nu}
\nu=\int_0^{s}\Psi_3^{(0)}\dd{s}
-\int_0^{s}\frac{\Pi_{+}\mathcal{Q}_+}{(\xi_{+})^2}\dd{s}
+c_{\nu}.
\end{equation}
The integrals used in the above expressions are
\begin{equation}
\label{Q}
\mathcal{Q}_{\pm}\equiv\int_0^{s}\bar{\Psi}_2^{(0)}\xi_{\pm}\dd{s}+c_{\mathcal{Q}}. 
\end{equation}
\begin{equation}
\label{Pi}
\Pi_{\pm}\equiv\int_0^{s}
\Psi_1^{(0)}\xi_{\pm}\dd{s}+c_{\pm},
\end{equation}
One can readily verify that this set of solutions satisfies the Sachs equations. This completes the determination of the spin coefficients. Finally, the quantities ${c_{\mathcal{Q}},c_{+},c_{-},c_{\gamma},c_{\nu}}$ are constants of integration. They are fixed by considering the asymptotic behavior of the spin coefficients near the emitter.

\subsection{Transport equations: The gradients of spin coefficients}
As discussed above, the gradients of the spin coefficients, e.g., $\bm{\delta}\rho$ and $\bm{\delta\delta}\rho$, cannot be calculated directly from their definitions. Using the commutator relations (\ref{commutators}), one can derive evolution equations for these gradients along the geodesics by acting with $\bm{\mathcal{D}}$. This set of equations, referred to as transport equations in Refs.\,\cite{Dolan_2018a,Dolan_2018b,Shipley_2019,Bruyere_2026}, is given by
\begin{subequations}
\label{SC-Grad}
\begin{align}
\bm{\mathcal{D}}(\bm{\delta}\rho)
&=-3\rho(\bm{\delta}\rho)
-\sigma(\bm{\delta}\rho)^*
-[\bar{\sigma}(\bm{\delta}\sigma)
+\sigma(\bar{\bm{\delta}}\sigma)^*]
-\tau(\rho^2+\sigma\bar{\sigma})\\
\bm{\mathcal{D}}(\bm{\delta}\sigma)
&=-3\rho(\bm{\delta}\sigma)
-\sigma(\bar{\bm{\delta}}\sigma)
-2\sigma(\bm{\delta}\rho)
-\tau(2\rho\sigma-\Psi_0)
+\bm{\delta}\Psi_0,\\
\bm{\mathcal{D}}(\bar{\bm{\delta}}\sigma)
&=-3\rho(\bar{\bm{\delta}}\sigma)
-\bar{\sigma}(\bm{\delta}\sigma)
-2\sigma(\bm{\delta}\rho)^*
-\bar{\tau}(2\rho\sigma-\Psi_0)
+\bar{\bm{\delta}}\Psi_0,\\
\bm{\mathcal{D}}(\bm{\delta}\alpha)
&=-2\rho(\bm{\delta}\alpha)
-\sigma(\bar{\bm{\delta}}\alpha)
-\bar{\sigma}(\bm{\delta}\beta)
-\alpha(\bm{\delta}\rho)
-\beta(\bar{\bm{\delta}}\sigma)^*
-\tau(\alpha\rho+\beta\bar{\sigma})\\
\bm{\mathcal{D}}(\bar{\bm{\delta}}\alpha)
&=-2\rho(\bar{\bm{\delta}}\alpha)
-\bar{\sigma}(\bm{\delta}\alpha)
-\bar{\sigma}(\bar{\bm{\delta}}\beta)
-\alpha(\bm{\delta}\rho)^*
-\beta(\bm{\delta}\sigma)^*
-\bar{\tau}
(\alpha\rho+\beta\bar{\sigma}),\\
\bm{\mathcal{D}}(\bm{\delta}\beta)
&=-2\rho(\bm{\delta}\beta)
-\sigma(\bar{\bm{\delta}}\beta)
-\sigma(\bm{\delta}\alpha)
-\beta(\bm{\delta}\rho)
-\alpha(\bm{\delta}\sigma)
-\tau
(\beta\rho+\alpha\sigma-\Psi_1)
+\bm{\delta}\Psi_1,\\
\bm{\mathcal{D}}(\bar{\bm{\delta}}\beta)
&=-2\rho(\bar{\bm{\delta}}\beta)
-\bar{\sigma}(\bm{\delta}\beta)
-\sigma(\bar{\bm{\delta}}\alpha)
-\beta(\bm{\delta}\rho)^*
-\alpha(\bar{\bm{\delta}}\sigma)
-\bar{\tau}
(\beta\rho+\alpha\sigma-\Psi_1)
+\bar{\bm{\delta}}\Psi_1,
\end{align}
\end{subequations}
and the second-order transport equations read
\begin{subequations}
\label{SC-Grad-Grad}
\begin{align}
&\begin{aligned}
\bm{\mathcal{D}}(\bm{\delta}\bm{\delta}\rho)
&=-4\rho(\bm{\delta}\bm{\delta}\rho)
-\sigma[(\bm{\delta}\bar{\bm{\delta}}\rho)
+(\bm{\delta}\bar{\bm{\delta}}\rho)^*]
-[\bar{\sigma}(\bm{\delta}\bm{\delta}\sigma)
+\sigma(\bar{\bm{\delta}}\bar{\bm{\delta}}\sigma)^*]\\
&\quad-3(\bm{\delta}\rho)^2
-(\bar{\bm{\delta}}\rho)(\bm{\delta}\sigma)
-2(\bm{\delta}\sigma)(\bar{\bm{\delta}}\sigma)^*
-(\bm{\delta}\tau)(\rho^2+\sigma\bar{\sigma})\\
&\quad-\tau[5\rho(\bm{\delta}\rho)
+\sigma(\bar{\bm{\delta}}\rho)
+2\sigma(\bar{\bm{\delta}}\sigma)^*
+2\bar{\sigma}(\bm{\delta}\sigma)]
-\tau^2(\rho^2+\sigma\bar{\sigma}),
\end{aligned}\\
&\begin{aligned}
\bm{\mathcal{D}}(\bm{\delta}\bar{\bm{\delta}}\rho)
&=-4\rho(\bm{\delta}\bar{\bm{\delta}}\rho)
-[\sigma(\bm{\delta}\bm{\delta}\rho)^*+\bar{\sigma}(\bm{\delta}\bm{\delta}\rho)]
-[\bar{\sigma}(\bm{\delta}\bar{\bm{\delta}}\sigma)
+\sigma(\bar{\bm{\delta}}\bm{\delta}\sigma)^*]\\
&\quad-3(\bm{\delta}\rho)(\bar{\bm{\delta}}\rho)
-(\bm{\delta}\rho)(\bar{\bm{\delta}}\sigma)^*
-[(\bm{\delta}\sigma)(\bm{\delta}\sigma)^*
+(\bar{\bm{\delta}}\sigma)(\bar{\bm{\delta}}\sigma)^*]
-(\bar{\bm{\delta}}\tau)^*(\rho^2+\sigma\bar{\sigma})\\
&\quad-\tau[3\rho(\bar{\bm{\delta}}\rho)
+\sigma(\bm{\delta}\sigma)^*
+\bar{\sigma}(\bar{\bm{\delta}}\sigma)]
-\bar{\tau}\left[(2\rho+\bar{\sigma})(\bm{\delta}\rho)
+\sigma(\bm{\delta}\sigma)^*
+\bar{\sigma}(\bar{\bm{\delta}}\sigma)\right]
-\tau\bar{\tau}(\rho^2+\sigma\bar{\sigma}),
\end{aligned}\\
&\begin{aligned}
\bm{\mathcal{D}}(\bm{\delta}\bm{\delta}\sigma)
&=-4\rho(\bm{\delta}\bm{\delta}\sigma)
-\sigma[(\bm{\delta}\bar{\bm{\delta}}\sigma)+(\bar{\bm{\delta}}\bm{\delta}\sigma)]
-2\sigma(\bm{\delta}\bm{\delta}\rho)
-5(\bm{\delta}\rho)(\bm{\delta}\sigma)
-(\bm{\delta}\sigma)(\bar{\bm{\delta}}\sigma)
-(\bm{\delta}\tau)(2\rho\sigma-\Psi_0)\\
&\quad-\tau\left[5\rho(\bm{\delta}\sigma)
+\sigma(\bar{\bm{\delta}}\sigma)
+4\sigma(\bm{\delta}\rho)\right]
-\tau^2(2\rho\sigma-\Psi_0)
+2\tau(\bm{\delta}\Psi_0)
+(\bm{\delta}\bm{\delta}\Psi_0),
\end{aligned}\\
&\begin{aligned}
\bm{\mathcal{D}}(\bm{\delta}\bar{\bm{\delta}}\sigma)
&=-4\rho(\bm{\delta}\bar{\bm{\delta}}\sigma)
-[\sigma(\bar{\bm{\delta}}\bar{\bm{\delta}}\sigma)
+\bar{\sigma}(\bm{\delta}\bm{\delta}\sigma)]
-2\sigma(\bm{\delta}\bar{\bm{\delta}}\rho)
-3(\bm{\delta}\rho)(\bar{\bm{\delta}}\sigma)
-2(\bar{\bm{\delta}}\rho)(\bm{\delta}\sigma)
-(\bm{\delta}\sigma)(\bar{\bm{\delta}}\sigma)^*
-(\bar{\bm{\delta}}\tau)^*(2\rho\sigma-\Psi_0)\\
&\quad-\tau\left[3\rho(\bar{\bm{\delta}}\sigma)
+\bar{\sigma}(\bm{\delta}\sigma)
+2\sigma(\bar{\bm{\delta}}\rho)\right]
-2\bar{\tau}[(\bm{\delta}\rho)\sigma
+\rho(\bm{\delta}\sigma)]
-\tau\bar{\tau}(2\rho\sigma-\Psi_0)
+\bar{\tau}(\bm{\delta}\Psi_0)
+\tau(\bar{\bm{\delta}}\Psi_0)
+(\bm{\delta}\bar{\bm{\delta}}\Psi_0),
\end{aligned}\\
&\begin{aligned}
\bm{\mathcal{D}}(\bar{\bm{\delta}}\bm{\delta}\sigma)
&=-4\rho(\bar{\bm{\delta}}\bm{\delta}\sigma)
-[\sigma(\bar{\bm{\delta}}\bar{\bm{\delta}}\sigma)
+\bar{\sigma}(\bm{\delta}\bm{\delta}\sigma)]
-2\sigma(\bm{\delta}\bar{\bm{\delta}}\rho)^*
-3(\bar{\bm{\delta}}\rho)(\bm{\delta}\sigma)
-2(\bm{\delta}\rho)(\bar{\bm{\delta}}\sigma)
-(\bar{\bm{\delta}}\sigma)^2
-(\bar{\bm{\delta}}\tau)(2\rho\sigma-\Psi_0)\\
&\quad-2\tau[\sigma(\bar{\bm{\delta}}\rho)+\rho(\bar{\bm{\delta}}\sigma)]
-\bar{\tau}\left[3\rho(\bm{\delta}\sigma)
+2\sigma(\bm{\delta}\rho)
+\sigma(\bar{\bm{\delta}}\sigma)\right]
-\tau\bar{\tau}(2\rho\sigma-\Psi_0)
+\bar{\tau}(\bm{\delta}\Psi_0)
+\tau(\bar{\bm{\delta}}\Psi_0)
+(\bar{\bm{\delta}}\bm{\delta}\Psi_0),
\end{aligned}\\
&\begin{aligned}
\bm{\mathcal{D}}(\bar{\bm{\delta}}\bar{\bm{\delta}}\sigma)
&=-4\rho(\bar{\bm{\delta}}\bar{\bm{\delta}}\sigma)
-\bar{\sigma}[(\bm{\delta}\bar{\bm{\delta}}\sigma)+(\bar{\bm{\delta}}\bm{\delta}\sigma)]
-2\sigma(\bm{\delta}\bm{\delta}\rho)^*
-5(\bm{\delta}\rho)^*(\bar{\bm{\delta}}\sigma)
-(\bm{\delta}\sigma)(\bm{\delta}\sigma)^*
-(\bm{\delta}\tau)^*(2\rho\sigma-\Psi_0)\\
&\quad-\bar{\tau}\left[5\rho(\bar{\bm{\delta}}\sigma)
+\bar{\sigma}(\bm{\delta}\sigma)
+4\sigma(\bm{\delta}\rho)^*\right]
-\bar{\tau}^2(2\rho\sigma-\Psi_0)
+2\bar{\tau}(\bar{\bm{\delta}}\Psi_0)
+(\bar{\bm{\delta}}\bar{\bm{\delta}}\Psi_0).
\end{aligned}
\end{align}
\end{subequations}
These two systems of transport equations are complete and solvable once the Weyl scalars (and their gradients) and the spin coefficients have been determined. The explicit expressions of Weyl scalars for Schwarzschild lensing are presented in Sec.\,\ref{sec:lensing}, and the calculations of their gradients are given in Appendix \ref{app:NP-formalism}.

\subsection{Transport equations: The gradients of GW amplitudes}
As mentioned at the beginning of this section, the third set of quantities, the gradients of the GW amplitudes, can also be obtained from two similar groups of transport equations. The first-order gradients, such as $\bm{\delta}A_{mm}$ and $\bm{\Delta}A_{mm}$, satisfy \cite{Shipley_2019}
\begin{subequations}
\label{A-Grad}
\begin{align}
&\bm{\mathcal{D}}(\bm{\delta}A)=-2\rho(\bm{\delta}A)
-\sigma(\bar{\bm{\delta}}A)
-[\rho\tau+(\bm{\delta}\rho)]A,\\
&\bm{\mathcal{D}}(\bar{\bm{\delta}}A)
=-2\rho(\bar{\bm{\delta}}A)
-\bar{\sigma}(\bm{\delta}A)
-[\rho\bar{\tau}+(\bm{\delta}\rho)^*]A.
\end{align}
\end{subequations}
and
\begin{equation}
\label{A-Grad-Delta}
\bm{\mathcal{D}}(\bm{\Delta}A)
=-\rho(\bm{\Delta}A)
-\bar{\tau}(\bm{\delta}A)
-\tau(\bar{\bm{\delta}}A)
-[\rho(\gamma+\bar{\gamma})+(\bm{\Delta}\rho)]A,
\end{equation}
where $\bm{\Delta}\rho$ can be directly calculated from the Sachs equation (\ref{Sachs-equation-Delta-rho}). Here, $A$ denotes an arbitrary amplitude component, either $A_{mm}$ or $A_{\bar{m}\bar{m}}$. Furthermore, the second-order gradients of the GW amplitudes satisfy \cite{Shipley_2019}
\begin{subequations}
\label{A-Grad-Grad}
\begin{align}
&\begin{aligned}
\bm{\mathcal{D}}(\bm{\delta}\bm{\delta}A)
&=-3\rho(\bm{\delta}\bm{\delta}A)
-\sigma(\bar{\bm{\delta}}\bm{\delta}
+\bm{\delta}\bar{\bm{\delta}})A
-3[(\bm{\delta}\rho)+\rho\tau](\bm{\delta}A)\\
&\qquad-[(\bm{\delta}\sigma)+\sigma\tau](\bar{\bm{\delta}}A)
-[(\bm{\delta}\bm{\delta}\rho)
+2\tau(\bm{\delta}\rho)
+\rho(\bm{\delta}\tau)
+\rho\tau^2]A,
\end{aligned}\\
&\begin{aligned}
\bm{\mathcal{D}}(\bm{\delta}\bar{\bm{\delta}}A)
&=-3\rho(\bm{\delta}\bar{\bm{\delta}}A)
-(\bar{\sigma}\bm{\delta}\bm{\delta}
+\sigma\bar{\bm{\delta}}\bar{\bm{\delta}})A
-2[(\bm{\delta}\rho)+\rho\tau](\bar{\bm{\delta}}A)\\
&\qquad-[(\bm{\delta}\rho)^*+(\bar{\bm{\delta}}\sigma)^*+\rho\bar{\tau}+\bar{\sigma}\tau](\bm{\delta}A)
-[(\bm{\delta}\bar{\bm{\delta}}\rho)
+\bar{\tau}(\bm{\delta}\rho)
+\tau(\bm{\delta}\rho)^*
+\rho(\bar{\bm{\delta}}\tau)^*
+\rho\tau\bar{\tau}]A,
\end{aligned}\\
&\begin{aligned}
\bm{\mathcal{D}}(\bar{\bm{\delta}}\bm{\delta}A)
&=-3\rho(\bar{\bm{\delta}}\bm{\delta}A)
-(\bar{\sigma}\bm{\delta}\bm{\delta}
+\sigma\bar{\bm{\delta}}\bar{\bm{\delta}})A
-2[(\bar{\bm{\delta}}\rho)+\rho\bar{\tau}](\bm{\delta}A)\\
&\qquad-[(\bm{\delta}\rho)+(\bar{\bm{\delta}}\sigma)+\rho\tau+\sigma\bar{\tau}](\bar{\bm{\delta}}A)
-[(\bar{\bm{\delta}}\bm{\delta}\rho)
+\bar{\tau}(\bm{\delta}\rho)
+\tau(\bm{\delta}\rho)^*
+\rho(\bar{\bm{\delta}}\tau)
+\rho\tau\bar{\tau}]A,
\end{aligned}\\
&\begin{aligned}
\bm{\mathcal{D}}(\bar{\bm{\delta}}\bar{\bm{\delta}}A)
&=-3\rho(\bar{\bm{\delta}}\bar{\bm{\delta}}A)
-\bar{\sigma}(\bar{\bm{\delta}}\bm{\delta}
+\bm{\delta}\bar{\bm{\delta}})A
-3[(\bar{\bm{\delta}}\rho)+\rho\bar{\tau}](\bar{\bm{\delta}}A)\\
&\qquad-[(\bar{\bm{\delta}}\sigma)+\bar{\sigma}\bar{\tau}](\bm{\delta}A)
-[(\bar{\bm{\delta}}\bar{\bm{\delta}}\rho)
+2\bar{\tau}(\bar{\bm{\delta}}\rho)
+\rho(\bm{\delta}\tau)^*
+\rho\bar{\tau}^2]A.
\end{aligned}
\end{align}
\end{subequations}
These sets of equations are also complete and solvable once the GW amplitudes, spin coefficients, and their gradients have been determined.

\section{\label{sec:lensing}Schwarzschild Lensing}

The preceding sections, Secs.\,\ref{sec:propagation-equations} and \ref{sec:auxiliary}, have established the theoretical framework for practical computation of GW amplitudes and polarizations. In this section, we apply this framework to the GW lensing by a Schwarzschild black hole.

\subsection{Geodesics}
In Schwarzschild coordinates ${t,r,\theta,\varphi}$, the line element of a Schwarzschild black hole of mass $M$ is \cite{Schwarzschild_1916}
\begin{equation}
\dd{s}^2=-f\dd{t}^2+\frac{1}{f}\dd{r}^2
+r^2(\dd\theta^2+\sin^2\theta\dd\varphi^2),
\end{equation}
where $f=1-2M/r$. The null geodesic 4-momentum takes the form \cite{Chandrasekhar1983,MTW,Wald_1984}
\begin{equation}
\label{null-geodesic-equations}
\bm{k}=\left\{\frac{1}{f},U_r,0,\frac{L}{r^2}\right\},
\end{equation}
where $k^{\mu}=\dd{x^{\mu}}/\dd{s}$ is the 4-momentum of the graviton, and $L$ is its orbital angular momentum. Here we have restricted the motion to the equatorial plane, $\theta=\pi/2$. The radial potential $U_r$ is
\begin{equation}
U_{r}\equiv\pm\sqrt{1-f\frac{L^2}{r^2}},
\end{equation}
The $\pm$ sign corresponds to the GW moving toward ($+$) or away from ($-$) the pericenter, respectively. Once the angular momentum $L$ is specified, the geodesic equations can be solved either numerically or analytically. The analytical solution in terms of elliptic integrals is reviewed in Appendix \ref{App-B}.

\subsection{NP tetrad and Weyl scalars}
The first leg of the null tetrad is given by Eq.\,(\ref{null-geodesic-equations}); the other three legs are constructed as \cite{Brodutch_2011,Vadapalli_2024}
\begin{equation}
\label{tetrad}
\bm{m}=\bm{m}_0+\zeta\bm{k},\quad
\bm{n}=\bm{n}_0+\zeta(\bm{m}_0+\bar{\bm{m}}_0)+\zeta^2\bm{k},
\end{equation}
where
\begin{equation}
\label{tetrad-n0}
\bm{n}_0=\frac{1}{2}\left\{1,-fU_r,0,-\frac{L}{r^2}f\right\}.
\end{equation}
and
\begin{equation}
\label{tetrad-m0}
\bm{m}_0=\frac{1}{\sqrt{2}}\left\{0,-\frac{L}{r}f,\frac{i}{r},\frac{U_{r}}{r}\right\},
\end{equation}
The undetermined function $\zeta(r)$, which ensures that the null tetrad is parallel-transported, is governed by \cite{Vadapalli_2024}
\begin{equation}
\label{zeta-eq}
\frac{\dd\zeta}{\dd{r}}=\frac{1}{\sqrt{2}}\frac{ML}{r^3}U_r^{-1}.
\end{equation}
The background Weyl scalars are
\begin{subequations}
\label{Weyl-background}
\begin{align}
&\Psi_0^{(0)}=\frac{3}{r^2}\frac{ML^2}{r^3},\\
&\Psi_1^{(0)}=\frac{3}{r^2}\left\{\frac{ML^2}{r^3}\zeta+\frac{1}{\sqrt{2}}\frac{ML}{r^2}U_r\right\},\\
&\Psi_2^{(0)}=\frac{3}{r^2}
\left\{\left[\frac{ML^2}{r^3}
+\frac{L^2}{r^2}\left(\zeta^2
-\frac{1}{2}\right)
+1\right]\frac{M}{r}
+\sqrt{2}\frac{ML}{r^2}\zeta U_{r}\right\},\\
&\Psi_3^{(0)}=\frac{3}{r^2}
\left\{\left[\frac{3ML^2}{r^3}
+\frac{L^2}{r^2}\left(\zeta^2-\frac{3}{2}\right)
+1\right]\frac{M}{r}\zeta
+\frac{1}{\sqrt{2}}\left[\frac{M}{r}
+\left(3\zeta^2
-\frac{1}{2}\right)\right]\frac{ML}{r^2}U_{r}\right\},\\
&\Psi_4^{(0)}=\frac{3}{r^2}
\left\{\left[\frac{M^2L^2}{r^4}
+\frac{ML^2}{r^3}(6\zeta^2-1)
+\frac{L^2}{r^2}\left(\zeta^4-3\zeta^2+\frac{1}{4}\right)
+2\zeta^2\right]\frac{M}{r}
+\sqrt{2}\left[\frac{2M}{r}
+(2\zeta^2-1)\right]\frac{ML}{r^2}\zeta U_{r}\right\}.
\end{align}
\end{subequations}
All background Weyl scalars in Schwarzschild spacetime are real, which leads to a series of simplifications in the evolution and transport equations. As with the null geodesic equations, Eq.\,(\ref{zeta-eq}) for $\zeta(r)$ can be solved either numerically or analytically once $L$ is specified. The analytical solution is reviewed in Appendix \ref{App-B}.

\subsection{\label{subsec:IV-setting}Initial conditions}
In this work, we assume a point-like GW source that is sufficiently distant from the lens. Near the emitter, the spin coefficients take the well-known asymptotic forms \cite{Chandrasekhar1983}
\begin{equation}
\label{spin-coefficients-emitter}
\sigma=\lambda=\tau=\gamma=\nu=0,\quad
\rho=\frac{1}{s},\quad
\mu=\frac{1}{2s},\quad
\alpha=\frac{\cot\psi}{2\sqrt{2}s},\quad
\beta=-\frac{\cot\psi}{2\sqrt{2}s}.
\end{equation}
$\psi$ is the inclination of the GW source. 

The asymptotic behavior of the GD vectors, $\xi_{\pm}\rightarrow s$, leads to $\rho\rightarrow1/s$ and $\sigma\rightarrow0$, which is consistent with Eq.\,(\ref{spin-coefficients-emitter}). Taking into account the asymptotic behavior of the Weyl scalars, $\Psi_n\rightarrow0$, we obtain $\mathcal{Q}_{\pm}\rightarrow c_{\mathcal{Q}}$ and $\Pi_{\pm}\rightarrow c_{\pm}$. From Eqs.\,(\ref{lambda}\,-\,\ref{nu}), we have
\begin{equation}
\begin{aligned}
\lambda\rightarrow0,\quad
\mu\rightarrow\frac{c_{\mathcal{Q}}}{s},\quad
\tau\rightarrow\frac{c_{+}}{s},\quad
\alpha\rightarrow\frac{1}{2}\left(\frac{c_{+}-c_{-}}{s}\right),\quad
\beta\rightarrow\frac{1}{2}\left(\frac{c_{+}+c_{-}}{s}\right),
\end{aligned}
\end{equation}
Comparing these expressions with Eq.\,(\ref{spin-coefficients-emitter}), we can determine the constants
\begin{equation}
c_{\mathcal{Q}}=\frac{1}{2},\quad c_{+}=0,\quad c_{-}=-\frac{1}{\sqrt{2}}\cot\psi.
\end{equation}
Finally, for $\gamma$ and $\nu$, we have $\gamma\rightarrow c_{\gamma}$ and $\nu\rightarrow c_{\nu}$, which fixes the remaining two constants:
\begin{equation}
c_{\gamma}=c_{\nu}=0.
\end{equation}
This completes the discussion of the initial conditions for the spin coefficients and fixes all integration constants. Additionally, evaluating the directional derivatives of Eq.\,(\ref{spin-coefficients-emitter}) yields the initial values of their gradients. All of them vanish except \cite{Bruyere_2026}
\begin{equation}
\bm{\delta}\alpha
=\bar{\bm{\delta}}\alpha
=-\bm{\delta}\beta
=-\bar{\bm{\delta}}\beta
=-\frac{1}{(2s\sin\psi)^2}.
\end{equation}

Consider a binary-like GW source sufficiently distant from the lens, and denote the GW amplitudes for the $+$ and $\times$ polarization modes at distance $s$ as $\mathcal{A}_{+}/s$ and $\mathcal{A}_{\times}/s$, respectively. The relative phase difference between the two polarization modes is taken to be $\pi/2$. Projecting the GW amplitude tensor and its gradients onto the NP tetrad, we find
\begin{equation}
\label{A-initial-value}
A_{mm}=\frac{\mathcal{A}_{+}+\mathcal{A}_{\times}}{s},\quad
A_{\bar{m}\bar{m}}=\frac{\mathcal{A}_{+}-\mathcal{A}_{\times}}{s},
\end{equation}
Clearly, in the vicinity of the emitter, the first-order gradients of $A_{(a)(b)}$ decay as $1/s^2$, and the second-order gradients decay as $1/s^3$. It is therefore natural to set their initial values to zero.

\subsection{\label{subsec:simplification}Simplification of the system of equations}
This work aims to trace the evolution of GW polarizations, at both leading and subleading orders, along the paths of gravitons. To achieve this, five coupled systems of equations have to be solved: Eqs.\,(\ref{Psi-LO}\,-\,\ref{Psi-4-SLO}) for the linearized Weyl scalars; Eqs.\,(\ref{evolution-equation-A-NP},\,\ref{evolution-equation-B-NP}) for the GW amplitudes; Eqs.\,(\ref{GDE-2-NP}\,-\,\ref{Pi}) for the GD vector and spin coefficients; Eqs.\,(\ref{SC-Grad},\,\ref{SC-Grad-Grad}) for the first- and second-order gradients of the spin coefficients; and Eqs.\,(\ref{A-Grad}\,-\,\ref{A-Grad-Grad}) for the first and second-order gradients of the LO GW amplitudes. Since all background Weyl scalars in Schwarzschild spacetime are real, the above systems are decoupled, reducible, and can be significantly simplified.

The simplifications are as follows. (1) The GD vector remains real everywhere, since its initial values are real. (2) All spin coefficients and their gradients are real for the same reason; e.g., $\sigma=\bar{\sigma}$ and $\bm{\delta}\sigma=(\bm{\delta}\sigma)^*$. Note, however, that $\bm{\delta}\sigma\neq\bar{\bm{\delta}}\sigma$, because the complex conjugate of $\bar{\bm{\delta}}\sigma$ is $\bm{\delta}\bar{\sigma}$, not $\bm{\delta}\sigma$. The above observations reduce Eqs.\,(\ref{SC-Grad},\,\ref{SC-Grad-Grad},\,\ref{A-Grad},\,\ref{A-Grad-Grad}) to simplified, decoupled forms. The resulting equations are presented as Eqs.\,(\ref{D-SC-Grad},\,\ref{D-SC-Grad-Grad},\,\ref{D-A-Grad},\,\ref{D-A-Grad-Grad}) in Appendix \ref{App-C}. (3) The gradients of the GW amplitudes satisfy $\bm{\delta}A=\bar{\bm{\delta}}A$, $\bm{\delta}\bm{\delta}A=\bar{\bm{\delta}}\bar{\bm{\delta}}A$, and $\bar{\bm{\delta}}\bm{\delta}A=\bm{\delta}\bar{\bm{\delta}}A$, which follow from the homogeneous form of Eqs.\,(\ref{D-mathscr-A-minus},\,\ref{D-mathscr-K-plus},\,\ref{D-mathscr-K-minus}), respectively. (4) Finally, the above considerations simplify the source terms of the SLO evolution equation (\ref{evolution-equation-B-NP}) to
\begin{equation}
i\mathcal{S}_{(a)(b)}
=\left[\bm{\delta}\bar{\bm{\delta}}A_{(a)(b)}\right]+(3\alpha+\beta)\left[\bm{\delta}A_{(a)(b)}\right]
+\left[(\bar{\bm{\delta}}\tau)-2\alpha^2+10\alpha\beta-4\beta^2-2\rho\mu-\lambda\sigma-3\Psi^{(0)}_2\right]A_{(a)(b)},
\end{equation}
Using Eq.\,(\ref{rho}), the formal solution for $B_{(a)(b)}$ is
\begin{equation}
\label{B-sol}
B_{(a)(b)}(s)=
-\frac{i}{\sqrt{\xi_{+}(s)\left|\xi_{-}(s)\right|}}
\int _{0}^{s}\mathcal{S}_{(a)(b)}(s') \sqrt{\xi_{+}(s')\left|\xi_{-}(s')\right|}\dd{s'}.
\end{equation}
With the equations thus simplified, we now turn to the numerical solution.

\section{\label{sec:numerical}Numerical Results}

We first summarize the computational procedure for obtaining the GW amplitudes and polarizations in the NP formalism. (1) Solve the null geodesic equation (\ref{null-geodesic-equations}) for a Schwarzschild black hole at a given impact parameter $b\equiv L/M$, either analytically (see Appendix \ref{App-B}) or numerically. This determines the GW trajectory. (2) Construct the null tetrad along each GW path using Eq.\,(\ref{tetrad}), and solve for $\zeta(r)$ from Eq.\,(\ref{zeta-eq}), either analytically (see Appendix \ref{App-B}) or numerically, and evaluate the background Weyl scalars via Eq.\,(\ref{Weyl-background}). (3) Solve Eq.\,(\ref{GDE-2-NP}) for the GD vector, and evaluate the spin coefficients, which govern the evolution of the LO amplitudes, from Eqs.\,(\ref{rho}\,-\,\ref{nu}). (4) Using the initial conditions (\ref{A-initial-value}), obtain the LO amplitudes by integrating Eq.\,(\ref{evolution-equation-A-NP}). (5) Substitute the spin coefficients into Eq.\,(\ref{SC-Grad}), impose the initial conditions of subsection \ref{subsec:IV-setting}, and use Eq.\,(\ref{Weyl-Grad}) to compute the gradients of the background Weyl scalars. Solving the resulting transport equations yields the first-order gradients of the spin coefficients. The same procedure, applied to Eqs.\,(\ref{SC-Grad-Grad}) and (\ref{Weyl-Grad-Grad}), gives the second-order gradients, in which Eq.\,(\ref{SC-Grad-Grad}) is numerically solved and Eq.\,(\ref{Weyl-Grad-Grad}) is applied. (6) Solve Eqs.\,(\ref{A-Grad},\,\ref{A-Grad-Delta},\,\ref{A-Grad-Grad}) for the gradients of the LO GW amplitudes. (7) All quantities needed to compute the source term $\mathcal{S}_{(a)(b)}$ and to solve the SLO evolution equation (\ref{evolution-equation-B-NP}) are now in hand. (8) Finally, compute the GW polarizations, the linearized Weyl scalars, using Eqs.\,(\ref{Psi-LO},\,\ref{Psi-2-SLO},\,\ref{Psi-3-SLO},\,\ref{Psi-4-SLO}).

Since the aforementioned calculation process involves many tedious but conceptually straightforward steps, only selected intermediate variables will be presented in the main text. Following our previous work \cite{Li_2025}, the lensing configuration is characterized by three parameters: the angular position of the source $\beta_{S}$, the source-lens distance $D_{LS}$, and the observer-lens distance $D_{L}$. The two distances are fixed at $D_{LS}=750\,M$ and $D_{L}=780\,M$, respectively. The angular position is normalized by the Einstein angle $\theta_{E}$ as $\beta_0=\beta_{S}/\theta_{E}$. In this work, $\beta_0$ takes four values: $\{0.2,\,0.4,\,0.6,\,0.8\}$. The inclination of the source is set to $\psi=\pi/6$ as a representative value. Under the geometric-optics approximation, a Schwarzschild lens produces two images of the background source (unless $\beta_0=0$): one is an upright image with even parity and positive magnification, and the other is an inverted image with odd parity and negative magnification \cite{Schneider_1992}. The parameters adopted in this work are listed in Table \ref{tab:lensing-parameters}. This table differs slightly from that in our previous work \cite{Li_2025}: a few symbol conventions have been modified, and some quantities have been added or removed as needed. All values have been recalculated using the analytical solution to the null geodesic equations (see Appendix \ref{App-B}), rather than relying on the approximate values obtained under the weak-deflection and thin-lens assumptions.

\renewcommand\arraystretch{1.4}
\begin{table}[h]
\centering
\caption{Example lensing parameters used in our calculations. The observer-lens and source-lens distances are fixed as $D_{L}=780M$ and $D_{LS}=750M$. The first two columns give the (normalized) source position. The remaining columns list the dimensionless impact parameter $b$, the affine parameters of the pericenter ($s_L$) and the caustic location ($s_c$), and the image magnification. For each $\beta_0$, the first row corresponds to the parity-even image, and the second row to the parity-odd image.}
\setlength{\tabcolsep}{5mm}{
\begin{tabular}{cccccc}
\hline\hline
$\beta_0\equiv\beta_{S}/\theta_E$ & 
$\beta_{S}\,({\mathrm{rad}})$ & 
$b$ & $s_L(M)$ &
$s_c(M)$ &
magnification \\ 
\hline
\multirow{2}{*}{$0.2$} & 
\multirow{2}{*}{$0.0100276$} & 
$44.5779041$ & 
$779.947483$ &
/ &
$+3.0301928$\\
\cline{3-6}
& & 
$37.0669895$ & 
$780.353340$ &
$1310.86965$ &
$-2.0460268$\\
\hline
\multirow{2}{*}{$0.4$} & 
\multirow{2}{*}{$0.0200552$} & 
$48.9204889$ & 
$780.172144$ &
/ &
$+1.8170024$ \\
\cline{3-6}
& & 
$33.8908840$ &
$780.995045$ &
$1174.72811$ &
$-0.83203758$ \\
\hline
\multirow{2}{*}{$0.6$} & 
\multirow{2}{*}{$0.0300828$} & 
$53.6403995$ & 
$780.674538$ &
/ &
$+1.4358105$\\
\cline{3-6}
& & 
$31.0777583$ & 
$781.936256$ &
$1084.38665$ &
$-0.44963640$\\
\hline
\multirow{2}{*}{$0.8$} & 
\multirow{2}{*}{$0.0401105$} & 
$58.7145114$ & 
$781.449701$ &
/ &
$+1.2608203$ \\
\cline{3-6}
& & 
$28.5999590$ & 
$783.181489$ &
$1021.80017$ &
$-0.27317099$\\
\hline\hline
\end{tabular}}

\label{tab:lensing-parameters}
\end{table}

A notable complication is the appearance of the caustic points inherent to the geometric-optics approximation \cite{Schneider_1992}, which persist even when the approximation is extended to higher order. Due to tidal forces exerted by the lens, the growing rate of the GD vector $\xi_{-}$ decreases or becomes negative after the GW passes through the pericenter. When $\xi_{-}$ reaches zero, the cross-sectional area of the geodesic congruence vanishes, and the GW amplitudes formally diverge. At this point, geometric optics breaks down, and wave effects dominate the behaviors of the lensed GW \cite{Li_2025}. Consequently, the evolution and transport equations become singular at the caustic. If the observer lies near a caustic, an extremely magnified GW image is expected \cite{Lo_2025}. When the observer is inside the caustic (i.e., $s_o<s_c$), the image has even parity, and the singularity does not affect the numerical integration of the evolution and transport equations. If, instead, the observer lies outside the caustic ($s_o>s_c$), the image is parity-odd, and the divergence must be handled explicitly. Unfortunately, in the case of $\beta_0<1$, the two images must have the opposite parities \cite{Schneider_1992}. Regularizing the singularities is therefore an unavoidable part of our numerical treatment. In the following text, we will take Eq.\,(\ref{evolution-equation-B-NP}) as an example to introduce our regularization strategy.

\begin{figure}[h]
\centering
\includegraphics[width=1.0\linewidth]{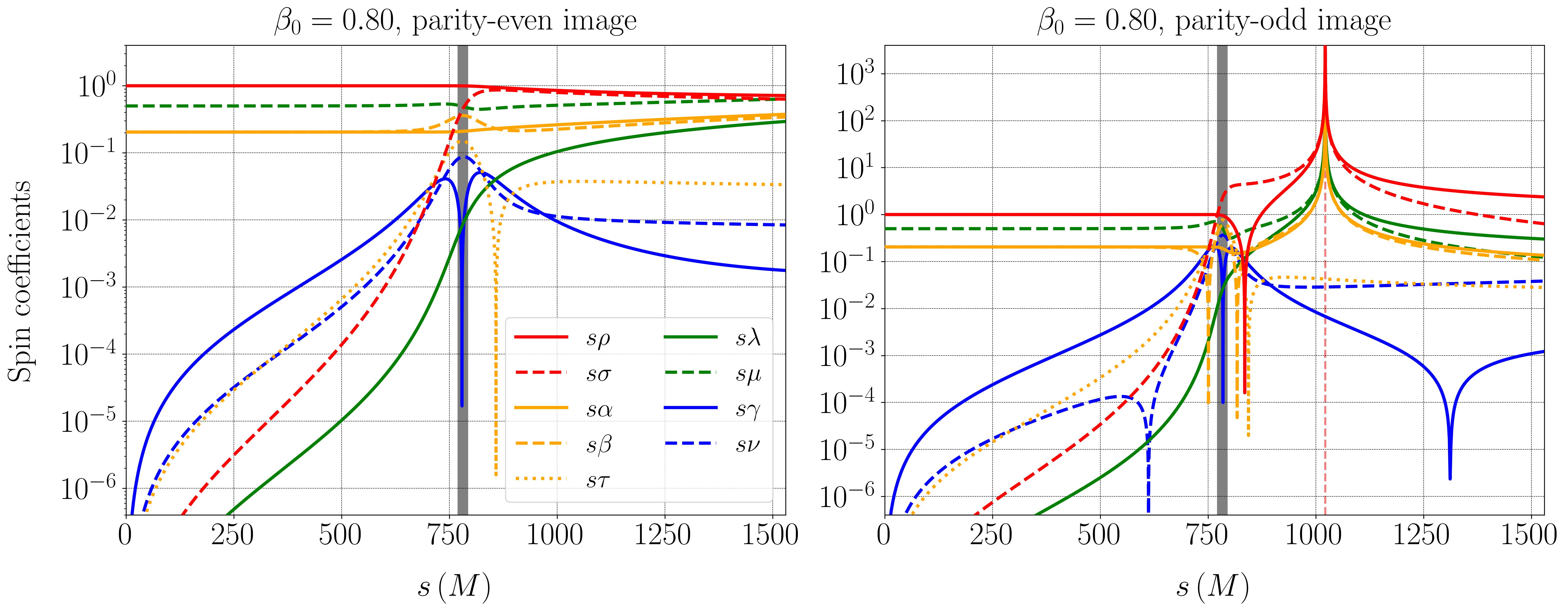}
\caption{Numerical solutions for the spin coefficients along the null geodesic for $\beta_0=0.8$, obtained from Eqs.\,(\ref{rho}\,-\,\ref{nu}). Different curves correspond to the various spin coefficients listed in the legend. The gray vertical lines mark the pericenter of the GW path, and the light coral dashed line indicates the caustic point (values given in Table \ref{tab:lensing-parameters}).}
\label{fig:spin-coefficients}
\end{figure}

Figure \ref{fig:spin-coefficients} illustrates the solutions of spin coefficients along the paths of the gravitons. As required, the tetrad is parallel-transported along the null geodesic, which forces three of the spin coefficients, $\kappa$, $\varepsilon$, and $\pi$, to vanish \cite{Chandrasekhar1983}. The remaining nine non-vanishing spin coefficients are all real numbers, as discussed in Sec.\,\ref{subsec:simplification}. In the left panel, which corresponds to parity-even images, the spin coefficients scaled by the affine parameter $s$ remain regular along the entire geodesic. In contrast, for the parity-odd case (right panel), six of the spin coefficients diverge at the caustic (marked by the light coral vertical line), whereas the remaining three stay regular. This can be understood from their analytical forms, Eqs.\,(\ref{tau},\,\ref{gamma},\,\ref{nu}), in which $\xi_{-}$ does not appear in the denominator.

\begin{figure}[h]
\centering
\includegraphics[width=1.0\linewidth]{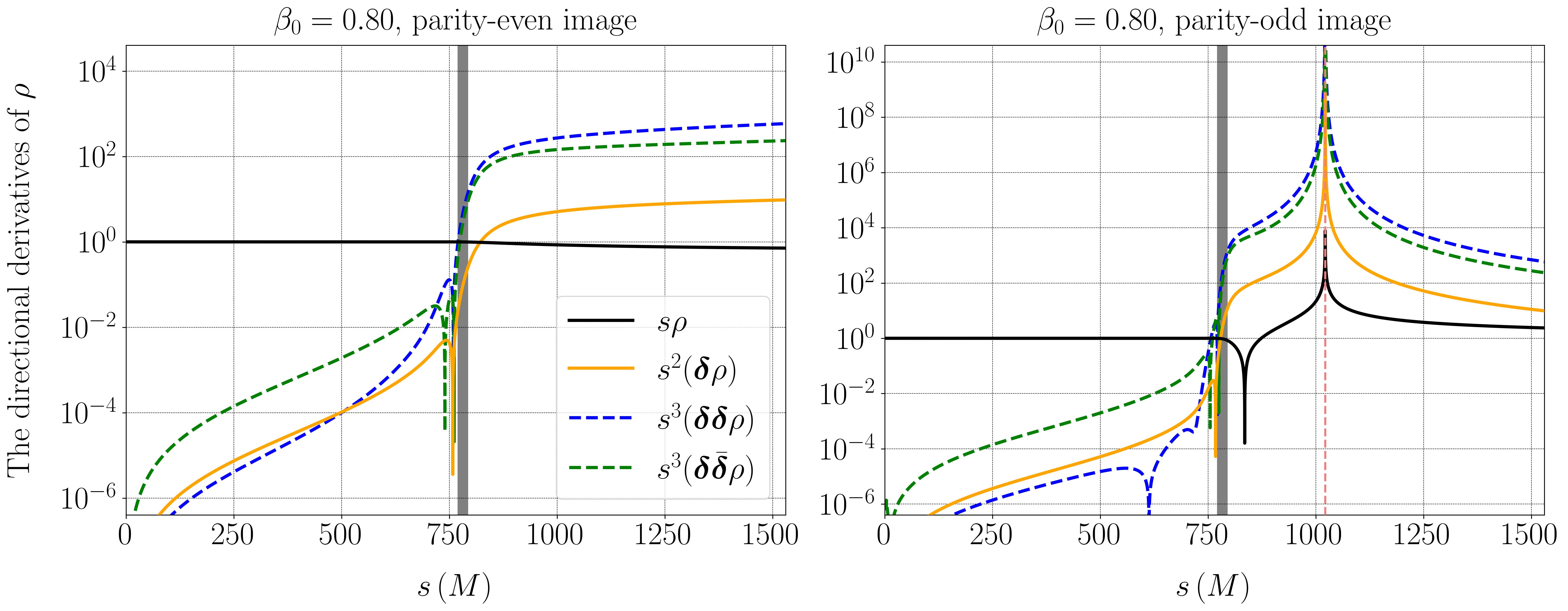}
\caption{Numerical solutions for $\rho$ (black solid) and its gradients $\bm{\delta}\rho$ (orange solid), $\bm{\delta}\bm{\delta}\rho$ (blue dashed), and $\bm{\delta}\bar{\bm{\delta}}\rho$ (green dashed) along the null geodesic for $\beta_0=0.8$, obtained from the simplified transport equations (\ref{D-SC-Grad},\,\ref{D-SC-Grad-Grad}), equivalent to the full system (\ref{SC-Grad},\,\ref{SC-Grad-Grad}). As in Fig.\,\ref{fig:spin-coefficients}, gray vertical lines mark the pericenter, and the light coral dashed line marks the caustic (see Table \ref{tab:lensing-parameters}).}
\label{fig:rho-grad}
\end{figure}

Figure \ref{fig:rho-grad} shows the corresponding results for the directional derivatives of the spin coefficients, using $\bm{\delta}\rho$, $\bm{\delta}\bm{\delta}\rho$, and $\bm{\delta}\bar{\bm{\delta}}\rho$ as representative examples. As noted earlier, because the GW phase $\Phi$ is a scalar, $\rho$ is real, and the spherical symmetry of the Schwarzschild background further forces $\bm{\delta}\rho=\bar{\bm{\delta}}\rho$. The second derivative $\bm{\delta}\bm{\delta}\rho$ is also real, but it is not equal to $\bm{\delta}\bar{\bm{\delta}}\rho$; instead, $\bm{\delta}\bm{\delta}\rho=\bar{\bm{\delta}}\bar{\bm{\delta}}\rho$. For the parity-odd image, caustic singularities are once again present.

\begin{figure}[h]
\centering
\includegraphics[width=0.85\linewidth]{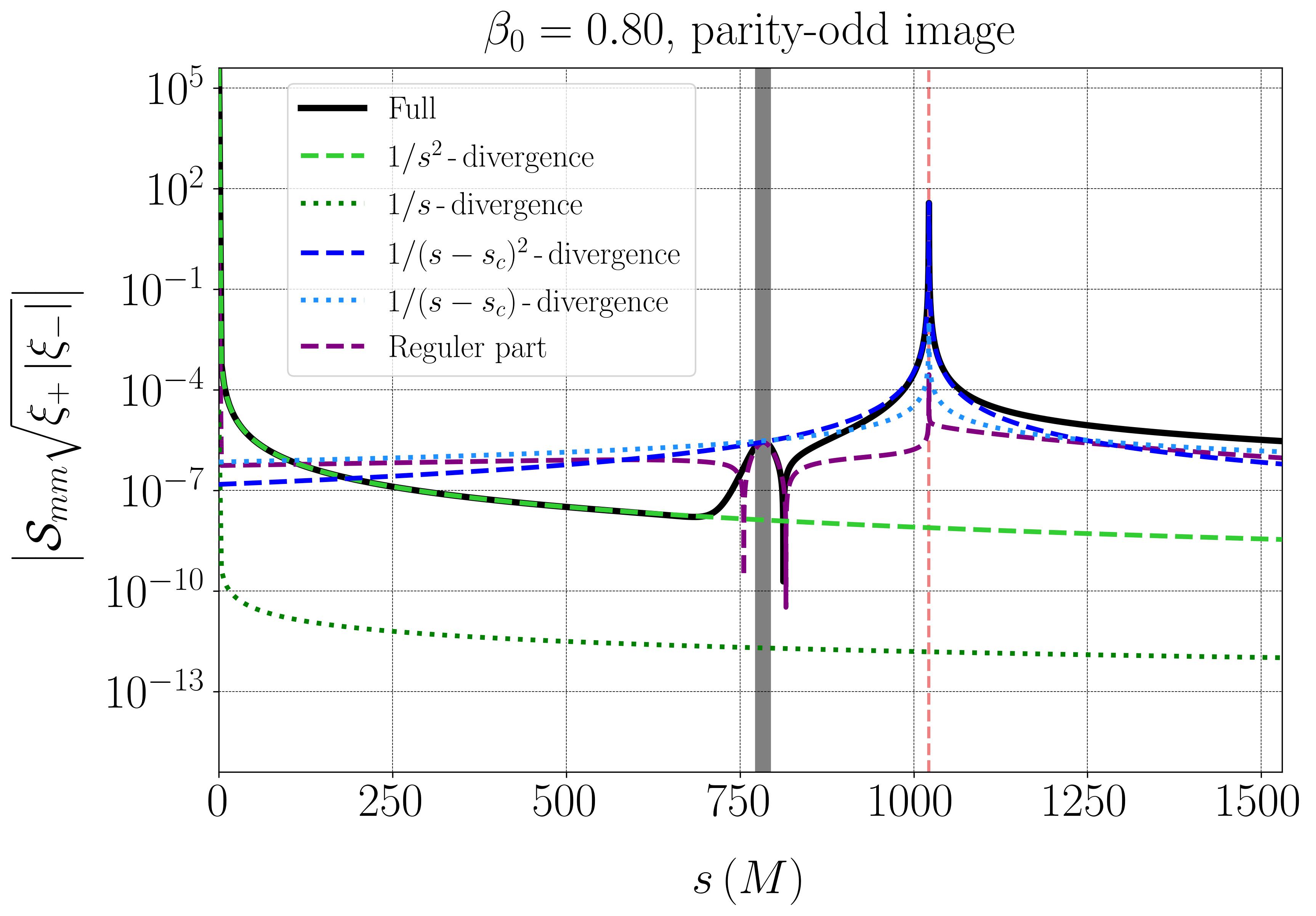}
\caption{Integrand of Eq.\,(\ref{B-sol}) with $\beta_0=0.80$ and odd parity. The integrand contains two singularities: one at the source position $s=0$ and the other at the caustic $s=s_c$. The figure shows the full integrand (black solid), its regular part (purple dashed), and the singular contributions, including the four divergence types $1/s$ (limegreen dashed), $1/s^2$ (green dotted), $1/(s-s_c)$ (blue dashed), and $1/(s-s_c)^2$ (dodgerblue dotted).}
\label{fig:source-B}
\end{figure}

Figure \ref{fig:source-B} illustrates the singularity and its regularization for the integrand of Eq.\,(\ref{B-sol}), $\mathcal{S}\equiv\mathcal{S}_{mm}\sqrt{\xi_{+}|\xi_{-}|}$. These singularities stem from $\xi_{-}$, which near $s_c$ behaves as $\xi_{-}\simeq\xi_{-}'(s_c)(s-s_c)$, where the prime denotes differentiation with respect to $s$. Consequently, physical quantities generically diverge as $\propto 1/(s-s_c)^n$. For $\mathcal{S}$, this singularity is of second order, that is, $\mathcal{S}\sim1/(s-s_c)^2$. To regularize this singularity, we define $$\alpha_2=\lim_{s\rightarrow s_{c}}(s-s_c)^2\mathcal{S},$$
and
$$\alpha_1=\lim_{s\rightarrow s_{c}}(s-s_c)\left[\mathcal{S}-\frac{\alpha_2}{(s-s_c)^2}\right],$$
such that $\widehat{\mathcal{S}}=\mathcal{S}-\alpha_1/(s-s_c)-\alpha_2/(s-s_c)^2$ is a regular function at $s_c$. Figure \ref{fig:source-B} displays the singular $\mathcal{S}$, its singular contributions $\alpha_2/(s-s_c)^2$ and $\alpha_1/(s-s_c)$, and the regular part $\widehat{\mathcal{S}}$. The integral of the regular part can then be evaluated numerically without difficulty, while the singular contributions are integrated analytically. Specifically, $\mathcal{S}$ not only has a singularity at $s_c$ but also at $s=0$, which is caused by the point-like wave-source assumption. This type of singularity can also be regularized using a similar method, which is not elaborated here.

\begin{figure}[h]
\centering
\includegraphics[width=0.9\linewidth]{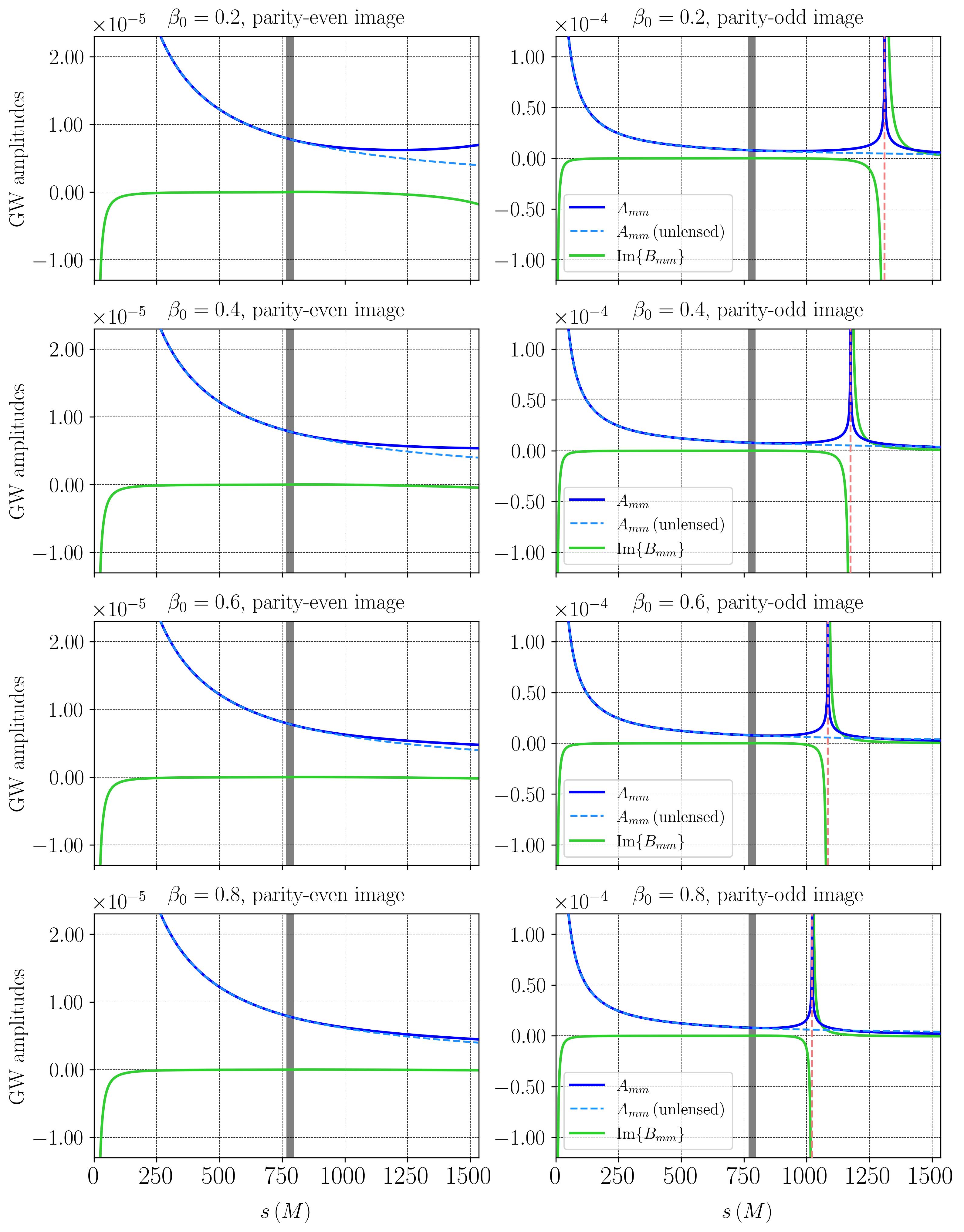}
\caption{Evolution of the GW amplitudes $A_{mm}$ and $B_{mm}$ along the GW paths, together with the unlensed $A_{mm}$ for comparison. The blue and dodgerblue dashed curves show the lensed and unlensed $A_{mm}$, respectively. The solutions for $B_{mm}$ are purely imaginary [see Eq.\,(\ref{B-sol})]; their imaginary parts are plotted as limegreen solid curves.}
\label{fig:Amp}
\end{figure}

Figure \ref{fig:Amp} shows the evolution of GW amplitudes $A_{mm}$ and $B_{mm}$ along the GW paths. Before the GW reaches the lens (denoted by the gray vertical lines), $A_{mm}$ decays as $1/s$, as shown by the overlap of the lensed (blue solid) and unlensed (light blue dashed) curves. In the same region, $B_{mm}$ decays as $1/s^2$, reflecting the fact that $B_{mm}$ does not contribute to the radiative DoF at the leading order. After passing through the pericenter, $A_{mm}$ deviates from the $1/s$ decay, resulting in a net amplification or de-amplification of the signal. Meanwhile, the SLO amplitudes $B_{mm}$ are excited by the GW-background interaction. Although $B_{mm}$ is generally much weaker than $A_{mm}$, it becomes appreciable for small source offsets, such as $\beta_0=0.2$ and $0.4$. Although $B_{mm}$ can also become large near the caustic, geometric optics breaks down in this region; quantitative predictions based on the geometric-optics approximation should be treated with caution there.

\begin{figure}[h]
\centering
\includegraphics[width=0.85\linewidth]{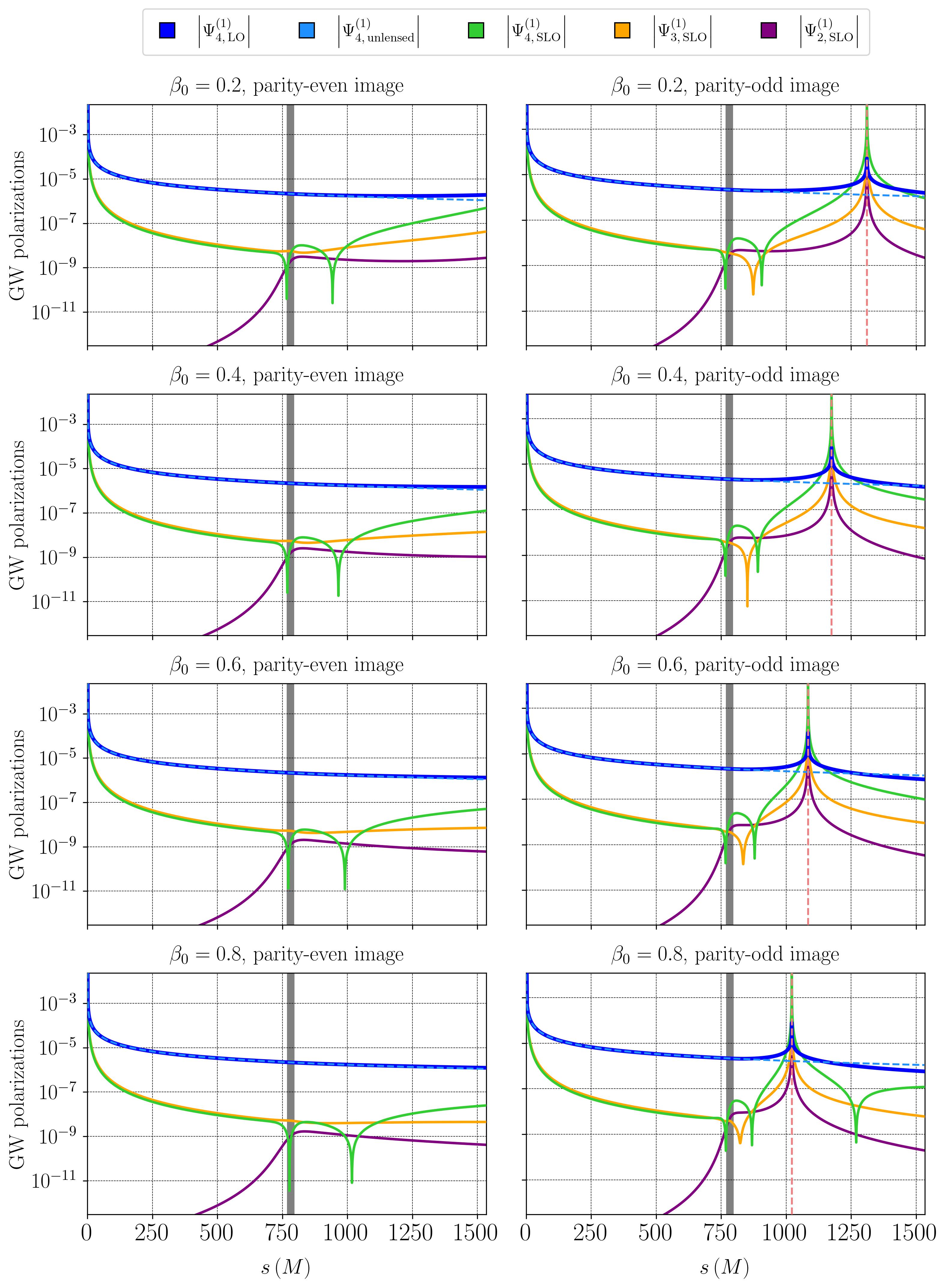}
\caption{Evolution of the linearized Weyl scalars along the GW paths, computed from Eqs.\,(\ref{Psi-LO}\,-\,\ref{Psi-4-SLO}). Blue solid curves: LO tensor mode $\Psi^{(1)}_{4,\mathrm{LO}}$; dodgerblue dashed curves: unlensed $\Psi^{(1)}_{4}$; limegreen curves: SLO tensor mode $\Psi^{(1)}_{4,\mathrm{SLO}}$; orange curves: vector mode $\Psi^{(1)}_{3,\mathrm{SLO}}$; purple curves: scalar mode $\Psi^{(1)}_{2,\mathrm{SLO}}$.}
\label{fig:Psi}
\end{figure}

Finally, Fig.\,\ref{fig:Psi} shows the evolution of the GW polarizations, i.e., the linearized Weyl scalars, as functions of the affine parameter $s$. For comparison, the unlensed $\Psi^{(1)}_{4}$ is also shown. As stated by Refs.\,\cite{Cusin_2020,Dalang_2022}, the lensing excites the apparent polarization modes, including both vector and scalar modes. Overall, the amplitudes decrease in the sequence: LO tensor mode [$\Psi^{(1)}_{4,\mathrm{LO}}$], SLO tensor mode [$\Psi^{(1)}_{4,\mathrm{SLO}}$], vector mode [$\Psi^{(1)}_{3,\mathrm{SLO}}$], and scalar mode [$\Psi^{(1)}_{2,\mathrm{SLO}}$]. A noteworthy feature is that $\Psi^{(1)}_{3,\mathrm{SLO}}$ does not always decay after the pericenter passage; for instance, it remains nearly constant for the parity-even image with $\beta_0=0.4$.


\section{\label{sec:conclusion}Conclusion and discussion}
The present work investigated the evolution of GW polarizations on curved backgrounds. The typical wavelength of the GW is assumed to be much smaller than the characteristic curvature scale of the background, $\lambdabar\ll\mathcal{R}$. We apply the short-wavelength expansion, using the small parameter $\epsilon\sim\lambdabar/\mathcal{R}$, to the linearized Einstein equation to analyze GW propagation \cite{MTW,Isaacson_1968}. At the leading order, the two polarization modes, the right-handed mode $A_{mm}$ and the left-handed mode $A_{\bar{m}\bar{m}}$, evolve independently along the null geodesics \cite{Hou_2019}. At the subleading order, the amplitude $B_{\mu\nu}$ is no longer orthogonal to the propagation direction, as follows from the SLO gauge conditions (see gauge condition, Eq.\,(\ref{gauge})). This implies that the polarization plane is effectively smeared and distorted by GW-background coupling, giving rise to apparent, unphysical polarization modes beyond the standard plus and cross modes \cite{Cusin_2020,Dalang_2022}. 

Using the NP formalism, we have systematically refined and improved the previous work. In Sec.\,\ref{sec:propagation-equations}, we discussed the gauge invariance of the Weyl scalars within the geometric-optics approximation and derived their explicit expressions entirely in terms of NP scalars \cite{Newman_Penrose_1962,Chandrasekhar1983} [see Eqs.\,(\ref{Psi-LO}\,-\,\ref{Psi-4-SLO})]. Subsequently, the equations governing the evolution of GW amplitudes [Eqs.\,(\ref{evolution-equation-A-NP},\,\ref{evolution-equation-B-NP})] were derived. Among them, $A_{(a)(b)}$ satisfies a first-order linear homogeneous ordinary differential equation, whose solution is completely determined by the initial conditions and the optical scalar $\rho$. In contrast, $B_{(a)(b)}$ satisfies an inhomogeneous equation, whose source term encodes the GW-background coupling.

To solve these evolution equations and evaluate the Weyl scalars, one is required to calculate a full set of spin coefficients, their gradients, and the gradients of the GW amplitudes. However, because these quantities are defined via transverse derivatives, their definitions cannot be applied directly in practical computations. To overcome this difficulty, an alternative method, built on GD, the Sachs equations, and transport equations, was systematically developed in Sec.\,\ref{sec:auxiliary}, following the approaches widely discussed in Refs.\,\cite{Gallo_2011,Boero_2019,Dolan_2018a}.

After constructing the full theoretical framework, we applied it to Schwarzschild lensing in Sec.\,\ref{sec:lensing} and presented the numerical results in Sec.\,\ref{sec:numerical}. Representative results for key intermediate quantities, such as the spin coefficients and their gradients, are shown in Figs.\,\ref{fig:spin-coefficients} and \ref{fig:rho-grad}, respectively. For the adopted parameters ($\beta_0<1$), Schwarzschild lensing produces two images with even and odd parities, respectively. For the parity-odd image, several quantities encounter caustic singularities, which necessitates appropriate regularization schemes when integrating the evolution and transport equations. Taking the integrand of Eq.\,(\ref{B-sol}) as an example, we show its divergent part and the regular part in Fig.\,\ref{fig:source-B}. Figure \ref{fig:Psi} shows the evolution of the Weyl scalars, the gauge-invariant characterization of GW polarizations, which is the central result of the detailed calculations performed in this work. Based on a more rigorous theoretical framework and detailed numerical solutions, we have confirmed the existence of apparent vector and scalar polarization modes, which originate from the smearing distortion of the wavefront induced by the GW-background coupling, and do not correspond to any new physical degrees of freedom.

This work presents a complete theoretical framework for studying the evolution of polarization modes of lensed GWs within a higher-order short-wavelength approximation. On the one hand, this framework fills the gap left by LO geometric optics \cite{Hou_2019} and the Kirchhoff diffraction integral \cite{Guo_2020}, both of which entirely lack polarization information, without incurring the prohibitive computational cost of full black hole scattering calculations \cite{Li_2025c}. This offers a theoretical cornerstone for providing more accurate waveform templates for GW lensing observations. The apparent polarization components found in our results could serve as a potential criterion for distinguishing lensed from unlensed GW signals. Future work should aim at extending these calculations to realistic astrophysical scales with improved numerical efficiency, and at performing a more rigorous assessment of the tetrad dependence of the apparent polarization modes.

\begin{acknowledgments}
We thank Xian Chen for the helpful discussion. This work is supported by the National Natural Science Foundation of China (Grant No.\,12473037), China Postdoctoral Science Foundation under Grant No.\,2025M783223, and the National Natural Science Foundation
of China (Grant No. 12325301).
\end{acknowledgments}

\appendix
\section{Newman-Penrose formalism}
\label{app:NP-formalism}

\subsection{Tetrad, derivative, and spin coefficients}
The tetrad formalism \cite{Newman_Penrose_1962,Chandrasekhar1983} is a powerful tool for investigating GW propagation and polarization. In this formalism, the basic geometric quantities of general relativity, the metric, connection, and Riemann tensor, are replaced by a null tetrad together with the spin coefficients, Weyl scalars, and Ricci scalars. The tetrad satisfies the orthogonality relation
\begin{equation}
\label{orthogonality}
\gamma_{\mu\nu}e_{(a)}^{\mu}e_{(b)}^{\nu}=\eta_{(a)(b)}
=\left(\begin{array}{cccc}
0 & -1 & 0 & 0\\
-1 & 0 & 0 & 0\\
0 & 0 & 0 & 1\\
0 & 0 & 1 & 0\\
\end{array}\right),
\end{equation}
and the spacetime metric can be expressed as
\begin{equation}
\gamma^{\mu\nu}=\eta^{(a)(b)}e_{(a)}^{\mu}e_{(b)}^{\nu}
\end{equation}
where
\begin{equation}
\eta^{(a)(b)}=\left[\eta_{(a)(b)}\right]^{-1}\quad\text{and}\quad
e^{(a)}_{\mu}=\left[e_{(a)}^{\mu}\right]^{-1}.
\end{equation}

In the tetrad formalism, partial derivatives with respect to the coordinates are projected onto the null tetrad, thereby defining the directional derivatives. For example, the directional derivative of a vector is
\begin{equation}
\label{directional-derivative-definition}
A_{(a),(b)}\equiv e_{(b)}^{\nu}\partial_{\nu}\left[e_{(a)}^{\mu}A_{\mu}\right].
\end{equation}
Similarly, the covariant derivative is projected onto the tetrad to give the intrinsic derivative,
\begin{equation}
\label{intrinsic-derivative-definition}
A_{(a)|(b)}\equiv e_{(a)}^{\mu}\left(\nabla_{\nu}A_{\mu}\right)e_{(b)}^{\nu},
\end{equation}
which is related to the directional derivatives by
\begin{equation}
\label{intrinsic-derivative}
A_{(a)|(b)}=A_{(a),(b)}-\eta^{(c)(d)}\gamma_{(c)(a)(b)}A_{(d)}.
\end{equation}
The coefficients $\gamma_{(c)(a)(b)}$ are Ricci rotation coefficients, defined as
\begin{equation}
\label{Ricci-rotation-corfficients}
\gamma_{(c)(a)(b)}=e^{\mu}_{(c)}
\left[\nabla_{\nu}e_{(a)\mu}\right]e_{(b)}^{\nu},
\end{equation}
where the first and second indices of $\gamma_{(c)(a)(b)}$ are antisymmetric.

The NP formalism employs a compact notation for the directional derivatives and spin coefficients. The 12 independent components of the Ricci rotation coefficients are denoted by the following symbols,
\begin{equation}
\label{spin-coefficient-definition}
\begin{aligned}
&\kappa=\gamma_{(3)(1)(1)},\quad
\rho=\gamma_{(3)(1)(4)},\quad
\varepsilon=\frac{1}{2}\left[\gamma_{(2)(1)(1)}+\gamma_{(3)(4)(1)}\right],\\
&\sigma=\gamma_{(3)(1)(3)},\quad
\mu=\gamma_{(2)(4)(3)},\quad
\gamma=\frac{1}{2}\left[\gamma_{(2)(1)(2)}+\gamma_{(3)(4)(2)}\right],\\
&\lambda=\gamma_{(2)(4)(4)},\quad
\tau=\gamma_{(3)(1)(2)},\quad
\alpha=\frac{1}{2}\left[\gamma_{(2)(1)(4)}+\gamma_{(3)(4)(4)}\right],\\
&\nu=\gamma_{(2)(4)(2)},\quad
\pi=\gamma_{(2)(4)(1)},\quad
\beta=\frac{1}{2}\left[\gamma_{(2)(1)(3)}+\gamma_{(3)(4)(3)}\right].
\end{aligned}
\end{equation}
which are collectively referred to as the spin coefficients. Exchanging indices $3$ and $4$ yields the complex conjugates of the spin coefficients, e.g., $\bar{\rho}$ and $\bar{\mu}$.

\subsection{Commutators}
Applying the commutator $[e_{(a)}^{\mu}\partial_{\mu},e_{(b)}^{\nu}\partial_{\nu}]$ to an arbitrary scalar field $\vartheta$ gives
\begin{equation}
\vartheta_{,(a)(b)}-\vartheta_{,(b)(a)}
=\eta^{(c)(d)}\left[\gamma_{(c)(b)(a)}
-\gamma_{(c)(a)(b)}\right]\left[\vartheta_{,(d)}\right].
\end{equation}
Expressing the Ricci rotation coefficients in terms of the spin coefficients leads to the following commutator relations,
\begin{subequations}
\label{commutators}
\begin{align}
{[\bm{\mathcal{D}},\bm{\Delta}]}
&=(\gamma+\bar{\gamma})\bm{\mathcal{D}}
-\bar{\tau}\bm{\delta}-\tau\bar{\bm{\delta}},\\
{[\bm{\mathcal{D}},\bm{\delta}]}
&=\tau\bm{\mathcal{D}}
-\rho\bm{\delta}-\sigma\bar{\bm{\delta}},\\
{[\bm{\mathcal{D}},\bar{\bm{\delta}}]}
&=\bar{\tau}\bm{\mathcal{D}}
-\rho\bar{\bm{\delta}}-\bar{\sigma}\bm{\delta}.
\end{align}
\end{subequations}

\subsection{Weyls scalars}
In the NP formalism, the spacetime geometry is encoded in the Ricci scalars and Weyl scalars, which are, respectively, the projections of the Ricci tensor and the Weyl tensor $C_{\mu\nu\alpha\beta}$ onto the null tetrad. In this work, the Ricci tensor has been set to zero due to the vacuum spacetime, such that the Weyl tensor is exactly the Riemann tensor, $C_{\mu\nu\alpha\beta}=R_{\mu\nu\alpha\beta}$, whether for the background or gravitational perturbation. The ten independent components of the Weyl tensor are encoded in five complex scalars, defined as
\begin{subequations}
\label{Weyl-scalars-definition}
\begin{align}
\Psi_0&\equiv-C_{(1)(3)(1)(3)},\\
\Psi_1&\equiv-C_{(1)(2)(1)(3)},\\
\Psi_2&\equiv-C_{(3)(1)(2)(4)},\\
\Psi_3&\equiv-C_{(2)(1)(2)(4)},\\
\Psi_4&\equiv-C_{(2)(4)(2)(4)},
\end{align}
\end{subequations}
with $C_{(a)(b)(c)(d)}\equiv C_{\mu\nu\alpha\beta}e^{\mu}_{(a)}e^{\nu}_{(b)}e^{\alpha}_{(c)}e^{\beta}_{(d)}$. Exchanging indices $(3)$ and $(4)$ gives the complex conjugates of the Weyl scalars, $\bar{\Psi}_0,\dots,\bar{\Psi}_4$.

\subsection{Sachs equations}
Sachs equations are derived from the Ricci identity, which is
\begin{equation}
\label{Ricci-identity}
[\nabla_{\alpha},\nabla_{\beta}]e_{(a)}^\mu
=-R^{\mu}_{\lambda\beta\alpha}e_{(a)}^\lambda,
\end{equation}
where $\nabla_{\alpha}$ and $R^{\alpha}_{\beta\mu\nu}$ are the covariant derivative operator and Riemann tensor compatible with the background. The commutator is defined as $[\nabla_{\alpha},\nabla_{\beta}]\equiv\nabla_{\alpha}\nabla_{\beta}-\nabla_{\beta}\nabla_{\alpha}$. Projecting Eq.\,(\ref{Ricci-identity}) onto the NP tetrad yields 18 identities \cite{Chandrasekhar1983},
\begin{subequations}
\label{Sachs-equations}
\begin{align}
&\bm{\mathcal{D}}\rho+\rho^2+\sigma\bar{\sigma}=0,\label{Sachs-equation-rho}\\
&\bm{\mathcal{D}}\sigma+2\rho\sigma=\Psi_0,\label{Sachs-equation-sigma}\\
&\bm{\mathcal{D}}\tau+\rho\tau+\sigma\bar{\tau}=\Psi_1,\label{Sachs-equation-tau}\\
&\bm{\mathcal{D}}\alpha+\alpha\rho+\beta\bar{\sigma}=0,\label{Sachs-equation-alpha}\\
&\bm{\mathcal{D}}\beta+\beta\rho+\alpha\sigma=\Psi_1,\label{Sachs-equation-beta}\\
&\bm{\mathcal{D}}\gamma+\alpha\tau+\beta\bar{\tau}=\Psi_2,\label{Sachs-equation-gamma}\\
&\bm{\mathcal{D}}\lambda+\lambda\rho+\mu\bar{\sigma}=0,\label{Sachs-equation-lambda}\\
&\bm{\mathcal{D}}\mu+\mu\rho+\lambda\sigma=\Psi_2,\label{Sachs-equation-mu}\\
&\bm{\mathcal{D}}\nu+\lambda\tau+\mu\bar{\tau}=\Psi_3,\label{Sachs-equation-nu}\\
&\bm{\Delta}\lambda-\bar{\bm{\delta}}\nu
-\lambda(\mu+\bar{\mu}+3\gamma-\bar{\gamma})
+\nu(3\alpha+\bar{\beta}-\bar{\tau})=-\Psi_4,\\
&\bm{\delta}\rho-\bar{\bm{\delta}}\sigma
+\rho(\bar{\alpha}+\beta)
-\sigma(3\alpha-\bar{\beta})=-\Psi_1,\\
&\bm{\delta}\alpha-\bar{\bm{\delta}}\beta+\mu\rho-\lambda\sigma+\alpha\bar{\alpha}+\beta\bar{\beta}-2\alpha\beta=-\Psi_2,\\
&\bm{\delta}\lambda-\bar{\bm{\delta}}\mu
+\mu(\alpha+\bar{\beta})
+\lambda(\bar{\alpha}-3\beta)=-\Psi_3,\\
&\bm{\delta}\nu-\bm{\Delta}\mu+\mu^2+\lambda\bar{\lambda}+\mu(\gamma+\bar{\gamma})
+\nu(\tau-\bar{\alpha}-3\beta)=0,\\
&\bm{\delta}\gamma-\bm{\Delta}\beta
+\gamma(\tau-\bar{\alpha}-\beta)-\beta(\gamma-\bar{\gamma}-\mu)+\alpha\bar{\lambda}-\nu\sigma+\mu\tau=0,\\
&\bm{\delta}\tau-\bm{\Delta}\sigma+(\mu\sigma+\bar{\lambda}\rho)
+\tau(\tau+\beta-\bar{\alpha})
-\sigma(3\gamma-\bar{\gamma})=0,\\
&\bm{\Delta}\rho-\bar{\bm{\delta}}\tau
-(\bar{\mu}\rho+\lambda\sigma)
+\tau(\bar{\beta}-\alpha-\bar{\tau})
+\rho(\gamma+\bar{\gamma})=-\Psi_2,\label{Sachs-equation-Delta-rho}\\
&\bm{\Delta}\alpha-\bar{\bm{\delta}}\gamma
+\nu\rho-\lambda(\tau+\beta)
+\alpha(\bar{\gamma}-\bar{\mu})
+\gamma(\bar{\beta}-\bar{\tau})=-\Psi_3.
\end{align}
\end{subequations}
The Sachs equations themselves provide a route for computing the spin coefficients. Unfortunately, this set of equations is nonlinear and difficult to decouple, which makes it rather hard to solve in practice. Refs.\,\cite{Newman_Penrose_1962,Pineault_1977} proposed a transformation that expresses $\rho$ and $\sigma$ via Eqs.\,(\ref{Sachs-equation-rho},\,\ref{Sachs-equation-sigma}), thereby linking them to the GD vector; this insight motivated later works to determine $\rho$ and $\sigma$ by solving the GD equation. The method was subsequently extended by Dolan \cite{Dolan_2018a,Dolan_2018b} to encompass all spin coefficients. In this work, we follow Dolan's approach to compute the spin coefficients for the Schwarzschild lensing system.

\subsection{Bianchi identities}
Projecting the Bianchi identity $R_{\alpha\beta[\gamma\delta;\lambda]}=0$ onto the NP tetrad yields $R_{(a)(b)[(c)(d)|(e)]}=0$, which is written as
\begin{equation}
\label{Bianchi-ID}
\begin{aligned}
&\bm{\mathcal{D}} \Psi_1
-\bar{\bm{\delta}} \Psi_0
-4\alpha\Psi_0+4\rho\Psi_1=0,\\
&\bm{\mathcal{D}}\Psi_2
-\bar{\bm{\delta}} \Psi_1
-\lambda \Psi_0
-2 \alpha \Psi_1
+3 \rho \Psi_2=0,\\
&\bm{\mathcal{D}}\Psi_3
-\bar{\bm{\delta}} \Psi_2
-2 \lambda \Psi_1
+2 \rho \Psi_3=0,\\
&\bm{\mathcal{D}}\Psi_4
-\bar{\bm{\delta}} \Psi_3
-3 \lambda \Psi_2
+2 \alpha \Psi_3
+\rho \Psi_4=0,\\
&\bm{\Delta}\Psi_0
-\bm{\delta} \Psi_1
+(4\gamma-\mu)\Psi_0
-2(\beta+2\tau)\Psi_1
+3 \sigma \Psi_2=0,\\
& \bm{\Delta}\Psi_1
-\bm{\delta} \Psi_2
+\nu \Psi_0
+2(\gamma-\mu)\Psi_1
-3 \tau \Psi_2
+2 \sigma \Psi_3=0.\\
&\bm{\Delta}\Psi_2
-\bm{\delta}\Psi_3
+2\nu\Psi_1
-3 \mu \Psi_2
+2(\beta-\tau)\Psi_3
+\sigma \Psi_4=0.\\
&\bm{\Delta}\Psi_3
-\bm{\delta}\Psi_4
+3\nu \Psi_2
-2 \gamma \Psi_3
-4 \mu \Psi_3
+(4\beta-\tau)\Psi_4=0.
\end{aligned}
\end{equation}

\subsection{\label{App-A:Weyl-gradient}Gradient of Weyl scalars}
By projecting the first-order covariant derivatives of the Weyl tensor onto the NP tetrad, we have the first-order gradients of the Weyl scalars:
\begin{subequations}
\label{Weyl-Grad}
\begin{align}
-\bm{\mathcal{D}}\Psi_0
&=(\nabla_{\lambda}C_{\mu\nu\alpha\beta})k^{\lambda}k^{\mu}m^{\nu}k^{\alpha}m^{\beta},\\
-\bm{\Delta}\Psi_0
&=(\nabla_{\lambda}C_{\mu\nu\alpha\beta})n^{\lambda}k^{\mu}m^{\nu}k^{\alpha}m^{\beta}
+4\gamma\Psi_0-4\tau\Psi_1,\\
-\bm{\delta}\Psi_0
&=(\nabla_{\lambda}C_{\mu\nu\alpha\beta})m^{\lambda}k^{\mu}m^{\nu}k^{\alpha}m^{\beta}
+4\beta\Psi_0-4\sigma\Psi_1,\\
-\bar{\bm{\delta}}\Psi_0
&=(\nabla_{\lambda}C_{\mu\nu\alpha\beta})\bar{m}^{\lambda}k^{\mu}m^{\nu}k^{\alpha}m^{\beta}
+4\alpha\Psi_0-4\rho\Psi_1.
\end{align}
\end{subequations}
Similarly, by projecting the second-order gradients of the Weyl tensor onto the NP tetrad, we have the expressions of the second-order gradients of the Weyl scalars, named as
\begin{subequations}
\label{Weyl-Grad-Grad}
\begin{align}
&\begin{aligned}
-\bm{\delta}\bm{\delta}\Psi_0
&=(\nabla_{\sigma}\nabla_{\lambda}C_{\mu\nu\alpha\beta})m^{\sigma}m^{\lambda}k^{\mu}m^{\nu}k^{\alpha}m^{\beta}
+4[(5\beta-\bar{\alpha})\beta
-(\gamma+\mu)\sigma+(\bm{\delta}\beta)]\Psi_0\\
&\quad+4[(\bar{\alpha}-7\beta+\tau)\sigma-\bm{\delta}\sigma]\Psi_1+12\sigma^2\Psi_2
+\bar{\lambda}(\bm{\mathcal{D}}\Psi_0)
-\sigma(\bm{\Delta}\Psi_0)
+(9\beta-\bar{\alpha})(\bm{\delta}\Psi_0)-8\sigma(\bm{\delta}\Psi_1),
\end{aligned}\\
&\begin{aligned}
-\bm{\delta}\bar{\bm{\delta}}\Psi_0
&=(\nabla_{\sigma}\nabla_{\lambda}C_{\mu\nu\alpha\beta})m^{\sigma}\bar{m}^{\lambda}k^{\mu}m^{\nu}k^{\alpha}m^{\beta}
+4[\alpha\bar{\alpha}
+3\alpha\beta
-\gamma\rho
-\lambda\sigma
+(\bm{\delta}\alpha)]\Psi_0
-4[2\rho\beta
+2\alpha\sigma
+(\bm{\delta}\rho)]\Psi_1\\
&\quad+12\rho\sigma\Psi_2
+\mu(\bm{\mathcal{D}}\Psi_0)
-\rho(\bm{\Delta}\Psi_0)
+4\alpha(\bm{\delta}\Psi_0)
+(\bar{\alpha}+3\beta)(\bar{\bm{\delta}}\Psi_0)
-4\rho(\bm{\delta}\Psi_1)
-4\sigma(\bar{\bm{\delta}} \Psi_1),\\
\end{aligned}\\
&\begin{aligned}
-\bar{\bm{\delta}}\bm{\delta}\Psi_0
&=(\nabla_{\sigma}\nabla_{\lambda}C_{\mu\nu\alpha\beta})\bar{m}^{\sigma}m^{\lambda}k^{\mu}m^{\nu}k^{\alpha}m^{\beta}
+4[5\alpha\beta-\beta\bar{\beta}
-\gamma\rho-\mu\rho
+(\bar{\bm{\delta}}\beta)]\Psi_0\\
&\quad+4[\rho(\tau-2\beta)
+(\bar{\beta}-5\alpha)\sigma
-(\bar{\bm{\delta}}\sigma)]\Psi_1
+12\rho\sigma\Psi_2\\
&\quad+\bar{\mu}(\bm{\mathcal{D}} \Psi_0)
-\rho(\bm{\Delta}\Psi_0)
+(5\alpha-\bar{\beta})(\bm{\delta}\Psi_0)
+4\beta(\bar{\bm{\delta}}\Psi_0)
-4\rho(\bm{\delta}\Psi_1)
-4\sigma(\bar{\bm{\delta}}\Psi_1),\\
\end{aligned}\\
&\begin{aligned}
-\bar{\bm{\delta}}\bar{\bm{\delta}}\Psi_0
&=(\nabla_{\sigma}\nabla_{\lambda}C_{\mu\nu\alpha\beta})\bar{m}^{\sigma}\bar{m}^{\lambda}k^{\mu}m^{\nu}k^{\alpha}m^{\beta}
+4[\alpha(3\alpha+\bar{\beta})
-\lambda\rho-\gamma\bar{\sigma}+(\bar{\bm{\delta}}\alpha)]\Psi_0\\
&\quad-4[\rho(5\alpha+\bar{\beta})-\bar{\sigma}\tau+(\bar{\bm{\delta}} \rho)]\Psi_1
+12\rho^2\Psi_2\\
&\quad
+\lambda(\bm{\mathcal{D}}\Psi_0)
-\bar{\sigma}(\bm{\Delta}\Psi_0)
+7 \alpha(\bar{\bm{\delta}} \Psi_0)
+\bar{\beta}(\bar{\bm{\delta}} \Psi_0)
-8\rho(\bar{\bm{\delta}}\Psi_1).
\end{aligned}
\end{align}
\end{subequations}

\section{Analytical solution to the geodesic equations}
\label{App-B}
The analytical solution of the geodesic equations can be expressed in terms of elliptic integrals \cite{NIST:DLMF}. When the GW propagates from infinity toward the pericenter, the affine parameter $s$ is given by
\begin{equation}
s=\tilde{s}(u)-\tilde{s}(u_s).
\end{equation}
On the outward branch from the pericenter to infinity, the affine parameter becomes
\begin{equation}
s=2\tilde{s}(u_2)-\tilde{s}(u_s)-\tilde{s}(u),
\end{equation}
where $u=M/r$, $u_s=M/r_s$, and $r_s$ is the radial coordinate of the source. The function $\tilde{s}(u)$ is
\begin{equation}
\label{solution-geodesic}
\begin{aligned}
\tilde{s}
&=\frac{M^2}{|L|}\frac{\sqrt{u_3-u_1}}{\sqrt{2}u_2u_3(u_2-u_1)}
\left\{\frac{(u_2-u_1)}{uu_1 \sqrt{u_3-u_1}}\sqrt{(u-u_1)(u-u_2)(u-u_3)}\right.\\
&\left.+\frac{1}{u_1^2}\left(\frac{u_2-u_1}{u_3-u_1}\right)
(u_1u_2+u_2u_3+u_1u_3)
\bm{\Pi}(\hat{\psi},k,\hat{\alpha})
-\frac{u_3(u_2-u_1)}{u_1(u_3-u_1)}\mathbf{F}(\hat{\psi},k)
-\left(\frac{u_1-u_2}{u_1}\right)\mathbf{E}(\hat{\psi},k)\right\},
\end{aligned}
\end{equation}
where $\mathbf{F}(\cdot,\cdot)$, $\mathbf{E}(\cdot,\cdot)$, and $\mathbf{\Pi}(\cdot,\cdot,\cdot)$ denote the incomplete elliptic integrals of the first, second, and third kind, respectively, and $u_1<0<u_2<u_3$ are the three roots of the cubic equation
\begin{equation}
2u^3-u^2+(M/L)^2=0.
\end{equation}
The quantities appearing in Eq.\,(\ref{solution-geodesic}) are defined as
\begin{equation}
\label{psi-k-alpha-hat-def}
\sin\hat{\psi}=\sqrt{\frac{u-u_1}{u_2-u_1}},\quad
k=\frac{u_2-u_1}{u_3-u_1},\quad\text{and}\quad
\hat{\alpha}=1-\frac{u_2}{u_1}.
\end{equation}

Similarly, the solution for $\zeta$ reads
\begin{equation}
\zeta=\text{sgn}(L)\Big\{\tilde{\zeta}(u)-\tilde{\zeta}(u_s)\Big\},
\end{equation}
on the inward branch, and
\begin{equation}
\zeta=\mathrm{sgn}(L)\Big\{2\tilde{\zeta}(u_2)-\tilde{\zeta}(u_s)-\tilde{\zeta}(u)\Big\},
\end{equation}
on the outward branch.
\begin{equation}
\tilde{\zeta}=\frac{u_1\sqrt{u_3-u_1}}{u_2-u_1}
\Bigg\{\frac{ku_3}{u_1}\mathbf{F}(\hat{\psi},k)
+\hat{\alpha}\mathbf{E}(\hat{\psi},k)\Bigg\},
\end{equation}
where $\hat{\psi}$, $k$, and $\hat{\alpha}$ are defined in Eqs.\,(\ref{psi-k-alpha-hat-def}).

\section{Simplified, decoupled form of the transport equations}
\label{App-C}
We first define the following auxiliary quantities:
\begin{equation}
\mathcal{P}_{n}^{(\pm)}\equiv\frac{1}{2}\left\{\left[\bm{\delta}\Psi^{(0)}_n\right]\pm\left[\bar{\bm{\delta}}\Psi^{(0)}_n\right]\right\},
\end{equation}
and
\begin{equation}
\mathcal{Q}_{\pm\pm\pm}\equiv
\frac{1}{2}\left\{\left[\bm{\delta}\bm{\delta}\Psi^{(0)}_0\right]
\pm\left[\bm{\delta}\bar{\bm{\delta}}\Psi^{(0)}_0\right]
\pm\left[\bar{\bm{\delta}}\bm{\delta}\Psi^{(0)}_0\right]
\pm\left[\bar{\bm{\delta}}\bar{\bm{\delta}}\Psi^{(0)}_0\right]\right\}.
\end{equation}
For the first-order gradients of the spin coefficients, we define
\begin{subequations}
\begin{align}
\mathscr{S}&\equiv(1/2)[(\bm{\delta}\sigma)-(\bar{\bm{\delta}}\sigma)],\\
\mathscr{R}_{\pm}&\equiv(\bm{\delta}\rho)\pm(1/2)[(\bm{\delta}\sigma)+(\bar{\bm{\delta}}\sigma)],\\
\mathscr{F}_{\pm}&\equiv(1/2)\left\{\left[(\bm{\delta}\alpha)+(\bar{\bm{\delta}}\alpha)\right]\pm\left[(\bm{\delta}\beta)+(\bar{\bm{\delta}}\beta)\right]\right\},\\
\mathscr{M}_{\pm}&\equiv(1/2)\left\{\left[(\bm{\delta}\alpha)-(\bar{\bm{\delta}}\alpha)\right]\pm\left[(\bm{\delta}\beta)-(\bar{\bm{\delta}}\beta)\right]\right\},
\end{align}
\end{subequations}
in terms of which the coupled system of equations reduces to
\begin{subequations}
\label{D-SC-Grad}
\begin{align}
\bm{\mathcal{D}}\mathscr{S}&=-(3\rho-\sigma)\mathscr{S}
+\mathcal{P}^{(0)}_{-},\\
\bm{\mathcal{D}}\mathscr{R}_{+}&=-3(\rho+\sigma)\mathscr{R}_{+}-\tau[(\rho+\sigma)^2-\Psi_0]+\mathcal{P}^{(0)}_{+},\\
\bm{\mathcal{D}}\mathscr{R}_{-}&=-(3\rho-\sigma)\mathscr{R}_{-}-\tau[(\rho-\sigma)^2+\Psi_0]-\mathcal{P}^{(0)}_{+},\\
\bm{\mathcal{D}}\mathscr{F}_{+}
&=-2(\rho+\sigma)\mathscr{F}_{+}
-\tau^2(\rho+\sigma)
-\tau(\mathscr{R}_{+}-\Psi_1)
+\mathcal{P}_{1}^{(+)},\\
\bm{\mathcal{D}}\mathscr{F}_{-}
&=-2\rho\mathscr{F}_{-}
+(\beta-\alpha)\mathscr{R}_{-}
+\tau[(\beta-\alpha)(\rho-\sigma)-\Psi_1]
-\mathcal{P}_{1}^{(+)},\\
\bm{\mathcal{D}}\mathscr{M}_{+}
&=-2\rho\mathscr{M}_{+}
+(\beta-\alpha)\mathscr{S}
+\mathcal{P}_{1}^{(-)},\\
\bm{\mathcal{D}}\mathscr{M}_{-}
&=-2(\rho-\sigma)\mathscr{M}_{-}+\tau\mathscr{S}
-\mathcal{P}_{1}^{(-)}.
\end{align}
\end{subequations}
For the second-order gradients of the spin coefficients, we define
\begin{subequations}
\begin{align}
\mathscr{U}_{\pm}&\equiv\Big[(\bm{\delta}\bm{\delta}\rho)+(\bm{\delta}\bar{\bm{\delta}}\rho)\Big]\pm(1/2)\Big[(\bm{\delta}\bm{\delta}\sigma)+(\bm{\delta}\bar{\bm{\delta}}\sigma)+(\bar{\bm{\delta}}\bm{\delta}\sigma)+(\bar{\bm{\delta}}\bar{\bm{\delta}}\sigma)\Big],\\
\mathscr{V}_{\pm}&\equiv\Big[(\bm{\delta}\bm{\delta}\rho)-(\bm{\delta}\bar{\bm{\delta}}\rho)\Big]\pm(1/2)\Big[(\bm{\delta}\bm{\delta}\sigma)-(\bm{\delta}\bar{\bm{\delta}}\sigma)-(\bar{\bm{\delta}}\bm{\delta}\sigma)+(\bar{\bm{\delta}}\bar{\bm{\delta}}\sigma)\Big],\\
\mathscr{W}_{\pm}&\equiv(1/2)\Big\{\Big[(\bm{\delta}\bm{\delta}\sigma)-(\bar{\bm{\delta}}\bar{\bm{\delta}}\sigma)\Big]\pm\Big[(\bar{\bm{\delta}}\bm{\delta}\sigma)-(\bm{\delta}\bar{\bm{\delta}}\sigma)\Big]\Big\},
\end{align}
\end{subequations}
and obtain the following decoupled equations:
\begin{subequations}
\label{D-SC-Grad-Grad}
\begin{align}
\bm{\mathcal{D}}\mathscr{U}_{+}
&=-4(\rho+\sigma)\mathscr{U}_{+}
-2\Big[3\mathscr{R}_{+}
+5\tau(\rho+\sigma)\Big]\mathscr{R}_{+}
-2\Big[(\rho+\sigma)^2-\Psi_0\Big](\tau^2+\mathscr{F}_{+})
+4\tau\mathcal{P}_0^{(+)}
+\mathcal{Q}_{+++},\\
\bm{\mathcal{D}}\mathscr{U}_{-}
&=-4\rho\mathscr{U}_{-}
-2\Big[\mathscr{R}_{+}+2\mathscr{R}_{-}
+\tau(5\rho-3\sigma)\Big]\mathscr{R}_{-}
-2\Big[(\rho-\sigma)^2+\Psi_0\Big](\tau^2+\mathscr{F}_{+})
-4\tau\mathcal{P}^{(+)}_{0}
-\mathcal{Q}_{+++},\\
\bm{\mathcal{D}}\mathscr{V}_{+}
&=-4\rho\mathscr{V}_{+}
-2\Big(\mathscr{R}_{+}-2\mathscr{S}\Big)\mathscr{S}
+2\Big[(\rho+\sigma)^2-\Psi_0\Big]\mathscr{M}_{-}
+\mathcal{Q}_{--+},\\
\bm{\mathcal{D}}\mathscr{V}_{-}
&=-4(\rho-\sigma)\mathscr{V}_{-}
-2\Big(\mathscr{R}_{-}-2\mathscr{S}\Big)\mathscr{S}
+2\Big[(\rho-\sigma)^2+\Psi_0\Big]\mathscr{M}_{-}
-\mathcal{Q}_{--+},\\
\bm{\mathcal{D}}\mathscr{W}_{+}
&=-4\rho\mathscr{W}_{+}
-2\Big[\mathscr{R}_{+}+2\mathscr{R}_{-}
+\tau(3\rho-\sigma)\Big]\mathscr{S}
+2\tau\mathcal{P}_0^{(-)}
+\mathcal{Q}_{-+-},\\
\bm{\mathcal{D}}\mathscr{W}_{-}
&=-4\rho\mathscr{W}_{-}
-2\Big[\mathscr{R}_{+}+\mathscr{R}_{-}
-\mathscr{S}+2\rho\tau\Big]\mathscr{S}
+2\tau\mathcal{P}_0^{(-)}
+\mathcal{Q}_{+--}.
\end{align}
\end{subequations}
For the first-order gradients of the GW amplitudes, we define
\begin{equation}
\mathscr{A}_{\pm}\equiv(1/2)\Big[(\bm{\delta}A)+(\bar{\bm{\delta}}A)\Big],
\end{equation}
which are governed by 
\begin{subequations}
\label{D-A-Grad}
\begin{align}
\bm{\mathcal{D}}\mathscr{A}_{+}
&=-(2\rho+\sigma)\mathscr{A}_{+}
-[(\bm{\delta}\rho)+\rho\tau]A,\\
\label{D-mathscr-A-minus}\bm{\mathcal{D}}\mathscr{A}_{-}
&=-(2\rho-\sigma)\mathscr{A}_{-}.
\end{align}
\end{subequations}
For the second-order gradients of the GW amplitudes, we define
\begin{subequations}
\begin{align}
\mathscr{J}_{\pm}&\equiv(1/2)
\Big\{\Big[(\bm{\delta}\bm{\delta}A)
+(\bar{\bm{\delta}}\bar{\bm{\delta}}A)\Big]
\pm\Big[(\bar{\bm{\delta}}\bm{\delta}A)
+(\bm{\delta}\bar{\bm{\delta}}A)\Big]\Big\},\\
\mathscr{K}_{\pm}&\equiv(1/2)
\Big\{\Big[(\bm{\delta}\bm{\delta}A)
-(\bar{\bm{\delta}}\bar{\bm{\delta}}A)\Big]
\pm\Big[(\bar{\bm{\delta}}\bm{\delta}A)
-(\bm{\delta}\bar{\bm{\delta}}A)\Big]\Big\},
\end{align}
\end{subequations}
satisfying
\begin{subequations}
\label{D-A-Grad-Grad}
\begin{align}
\bm{\mathcal{D}}\mathscr{J}_{+}
&=-(3\rho+2\sigma)\mathscr{J}_{+}
-2\Big[2\mathscr{R}_{+}+\mathscr{R}_{-}+\tau(3\rho+\sigma)\Big]\mathscr{A}_{+}
-\frac{1}{2}\Big[4\rho\mathscr{F}_{+}
+\mathscr{U}_{+}+\mathscr{U}_{-}
+4\tau\left(\mathscr{R}_{+}+\mathscr{R}_{-}+\rho \tau\right)\Big]A\\
\bm{\mathcal{D}}\mathscr{J}_{-}
&=-(3\rho-2\sigma)\mathscr{J}_{-}
-2\mathscr{S}\mathscr{A}_{+}
-\frac{1}{2}\left(\mathscr{V}_{+}+\mathscr{V}_{-}-4\rho\mathscr{M}_{-}\right)A\\
\label{D-mathscr-K-plus}
\bm{\mathcal{D}}\mathscr{K}_{+}
&=-3\rho\mathscr{K}_{+}
-\Big[\mathscr{R}_{+}+3\mathscr{R}_{-}+2(2\rho-\sigma)\tau\Big]\mathscr{A}_{-},\\
\label{D-mathscr-K-minus}
\bm{\mathcal{D}}\mathscr{K}_{-}
&=-3\rho\mathscr{K}_{-}
-\Big[\mathscr{R}_{+}+\mathscr{R}_{-}-2(\mathscr{S}-\rho\tau)\Big]\mathscr{A}_{-}.
\end{align}
\end{subequations}

\bibliographystyle{apsrev4-2}
\bibliography{reference}

\end{document}